\documentclass[a4paper, amsfonts, amssymb, amsmath, reprint, showkeys, nofootinbib, twoside,superscriptaddress]{revtex4-1}
\usepackage{xcolor}
\usepackage{hyperref}
\usepackage{graphicx}
\usepackage{subcaption}
\usepackage{wrapfig}
\usepackage{float}
\usepackage{dcolumn}
\usepackage{bm}

\newcommand{\R} {{\mathbb R}}

\newcommand{\nosx} {{\overline{x}}}
\newcommand{\om} {{\omega}}

\newcommand{\Av}[2] {\left \langle  {#2} \right \rangle_{#1}}
\newcommand{\centr} {{0}}

\begin{document}

\preprint{APS/123-QED}

\title{
Riding Wavelets: A Method to Discover New Classes of Price Jumps
}

\author{Cécilia Aubrun}
 \altaffiliation[]{Authors contributed equally.}
\affiliation{Chair of Econophysics and Complex Systems, \'Ecole polytechnique, 91128 Palaiseau Cedex, France}
\affiliation{LadHyX UMR CNRS 7646, \'Ecole polytechnique, 91128 Palaiseau Cedex, France}

\author{Rudy Morel}
\altaffiliation[]{Authors contributed equally.}
\affiliation{École Normale Supérieure, 45 rue d'Ulm, 75005 Paris}
\affiliation{Chair of Econophysics and Complex Systems, \'Ecole polytechnique, 91128 Palaiseau Cedex, France}

\author{Michael Benzaquen}
\affiliation{Chair of Econophysics and Complex Systems, \'Ecole polytechnique, 91128 Palaiseau Cedex, France}
\affiliation{LadHyX UMR CNRS 7646, \'Ecole polytechnique, 91128 Palaiseau Cedex, France}
\affiliation{Capital Fund Management, 23 Rue de l’Universit\'e, 75007 Paris, France}

\author{Jean-Philippe Bouchaud}
\affiliation{Chair of Econophysics and Complex Systems, \'Ecole polytechnique, 91128 Palaiseau Cedex, France}
\affiliation{Capital Fund Management, 23 Rue de l’Universit\'e, 75007 Paris, France}
\affiliation{Académie des Sciences, 23 Quai de Conti, 75006 Paris, France\smallskip}

\date{\today}

\begin{abstract}
Cascades of events and extreme occurrences have garnered significant attention across diverse domains such as financial markets, seismology, and social physics.  
Such events can stem either from the internal dynamics inherent to the system (endogenous), or from external shocks (exogenous). The possibility of separating these two classes of events has critical implications for professionals in those fields. We introduce an unsupervised framework leveraging a representation of jump time-series based on wavelet coefficients and apply it to stock price jumps. In line with previous work, we recover the fact that the time-asymmetry of volatility is a major feature. Mean-reversion and trend are found to be two additional key features, allowing us to identify new classes of jumps. Furthermore, thanks to our wavelet-based representation, we investigate the reflexive properties of co-jumps, which occur when multiple stocks experience price jumps within the same minute. We argue that a significant fraction of co-jumps results from an endogenous contagion mechanism.
\end{abstract}

\maketitle




\section*{Introduction}



Extreme events and cascades of events are widespread occurrences in both natural and social systems \cite{sornette2006endogenous}. Examples include earthquakes, volcanic eruptions, hurricanes, epileptic crises \cite{osorio2010epileptic,sornette2010prediction}, epidemic spread, financial crashes \cite{filimonov2012quantifying,hardiman2013critical,hardiman2014branching}, economic crises \cite{bak1993aggregate, moran2019may}, book sales shocks \cite{sornette2004endogenous,deschatres2005dynamics}, riot propagation \cite{mohler2011self,bonnasse2018epidemiological} or failures in socio-technical systems \cite{moran2023temporal}.
Understanding the origin of such events is essential for forecasting and possibly stabilizing their dynamics.

A widely studied question is the reflexive nature of those shocks -- the concept of financial market reflexivity was introduced by Soros in \cite{sorosalchemy}, to describe the idea that price dynamics are mostly endogenous and arise from feedback mechanisms.  
Extreme events, in particular, are considered to be endogenous when they arise from feedback mechanisms within the system's structure ~\cite{bak1996nature,bak1995complexity, sornette2006endogenous}. Quantifying the extent of reflexivity in a complex system and distinguishing events caused by external shocks from those provoked endogenously, and more generally identifying different classes of events, are crucial questions. 

Prior research  has proposed to differentiate between endogenous and exogenous dynamics by analyzing the profile of activity around the shock \cite{sornette2003endogenous,sornette2004endogenous,deschatres2005dynamics,crane2008robust}, in particular in the context of financial markets \cite{vol_news_jp,marcaccioli2022exogenous}. 
It has been observed that endogenous shocks are preceded by a growth phase mirroring the post event power-law relaxation, in contrast to exogenous shocks that are strongly asymmetric. The universality of this result is quite intriguing as they have been observed in various contexts: intraday book sales on Amazon \cite{sornette2004endogenous,deschatres2005dynamics}, daily views of YouTube videos \cite{crane2008robust} and intraday financial market volatility and price jumps \cite{vol_news_jp, marcaccioli2022exogenous}. Besides, Wu \textit{et al.} \cite{wu2022classification} differentiate exogenous and endogenous bursts of comment posting on social media using the analysis of collective emotion dynamics and time-series distributions of comment arrivals. 
%

\begin{figure}
\centering
\includegraphics[width=\linewidth]{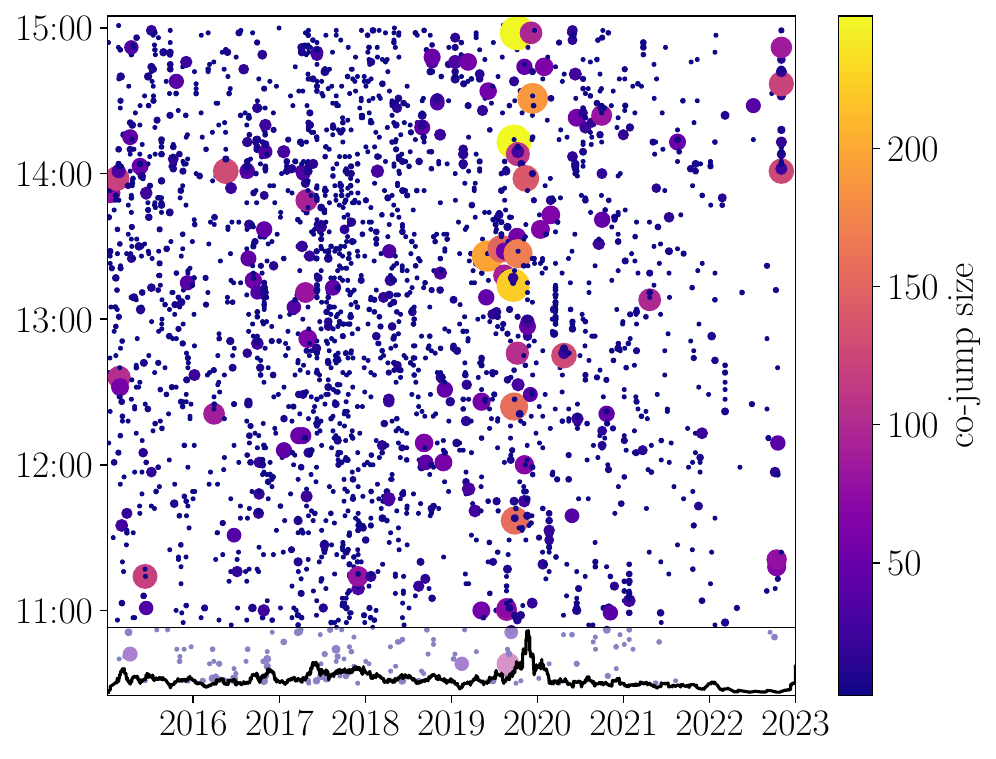}
\caption{
Visualization of our co-jumps dataset (295 US stocks, 8 years) (as in \cite{bormetti2015modelling,calcagnile2018collective, aubrun2023multivariate}). 
The horizontal axis corresponds to the day of the co-jump and the vertical axis gives the time of day. 
The size and color of the circle
encode the number of stocks jumping simultaneously (in the same minute). 
Inset: number of jumps on a rolling window of 30 days.
}
\label{fig:cojumpsillu}
\end{figure}

Furthermore, in complex systems, occurrences can propagate along two directions: temporally and towards the other elements of the system. Financial markets offer an attractive setting for studying multi-dimensional shocks due to the abundance of available data, the frequent occurrence of financial shocks and price jumps and the inter-connectivity of markets.
In fact, a recent study by Lillo \textit{et al.}~\cite{bormetti2015modelling, calcagnile2018collective} demonstrates the frequent occurrence of ``co-jumps'', defined as simultaneous jumps of multiple stocks (as illustrated in Fig.~\ref{fig:cojumpsillu}) and establishes a correlation between their prevalence and the inter-connectivity of different markets.

In this paper, we address the problem of classifying financial price jumps (and co-jumps), in particular measuring their reflexivity, by analyzing their time-series using wavelets. We introduce an unsupervised classification based on an embedding $\Phi(x)$ of each jump time-series of returns $x(t)$ into a low dimensional-space more appropriate to clustering. 
Such embedding, composed of wavelet scattering coefficients (see~\cite{bruna2013invariant} and below), relies on wavelet coefficients $Wx(0)$ of the time-series at the time of the jump $t=0$ and wavelet coefficients of volatility $W|Wx|(0)$.  
Such coefficients are particularly suitable to characterize (among other properties) the asymmetry of time-series at multiple scales. 

Through a Principal Component Analysis we retrieve the fact that time-asymmetry of volatility indeed plays an important role for classification. 
However, our analysis identifies two further crucial features for characterizing the nature of price jumps: mean-reversion and trend. Specifically, mean-reverting jumps are such that pre-jump and post-jump returns are of opposite signs, whereas trend-aligned and trend-anti-aligned jumps occur on a sequence of returns of same sign before and after the jump, but either aligned with the jump itself, or of opposite sign.

For each jump, our analysis provides a measure of the volatility asymmetry, the mean-reversion and the trend. 
We propose a visualization of our dataset of price jumps in the form of {\it two} 2D projections.
For both projections, one direction characterizes price jumps based on volatility asymmetry, or ``reflexivity level''. The second direction characterizes jumps either in terms of mean-reversion, or in terms of alignement with the local trend behavior. One can then measure the endogeneity of price co-jumps, revealing that {many jumps/co-jumps are {\it not} related to news and arise only due to endogenous dynamics.} This is consistent with the observed power-law distribution of the number of firms affected by a co-jump, indeed predicted by a simple branching (or contagion) process.

Surprisingly, we uncover that a significant number of large co-jumps (affecting a large number of stocks), which might have been assumed to be caused by a common factor and thus share analogous dynamics, actually have uncorrelated returns both pre- and post-jump. This again suggests that such jumps are mostly of endogenous origin.  

The outline of our paper is as follows.
Section~\ref{sec:SecData} describes our dataset of price jumps resulting from Marcaccioli \textit{et al.}~\cite{marcaccioli2022exogenous}, reviews their supervised classification method based on news labels, and investigates its limitation. 
Section \ref{sec:SecUnivariateJumpClassif} presents our unsupervised classification of univariate jump time-series based on wavelet coefficients.
Such classification identifies three main directions in the dataset, the time-asymmetry, the mean-reversion and the trend.  
Finally, section \ref{sec:cojumps} 
is devoted to the characterization of the endogeneity of co-jumps.

\section{\label{sec:SecData} Supervised classification through reflexivity}

Prior work has identified \textit{reflexivity} as an important feature for the classification of jumps in financial markets~\cite{marcaccioli2022exogenous}.
Given the time-series of a jump, the main challenge is to efficiently measure such reflexivity.  


One can for example look at contemporaneous news labels to determine whether a jump is exogenous. Indeed, news labels may serve as ground truth to learn a classification model on the activity profile around a shock. To exemplify, Fig.~\ref{fig:sym_asym_examples}, from the work of Marcaccioli \textit{et al.} \cite{marcaccioli2022exogenous}, illustrates the time asymmetry difference between endogenous and exogenous jumps.

In this section, we first introduce the jump detection method, which allows us to build our dataset. Then, we present the supervised classification based on news labels introduced in~\cite{marcaccioli2022exogenous} and show its limitations. This will motivate an alternative approach in section \ref{sec:SecUnivariateJumpClassif}.

\subsection{\label{sec:jumpDetection}Jump detection}

We refer to \cite{vol_news_jp, marcaccioli2022exogenous,boudt2011robust} for a detailed description of the method to detect price jumps. The detection relies on an estimator of ``jump-score" $x(t) = {r(t)}/{(f(t)\sigma(t))}$, which is the ratio of 1-minute returns time-series $r(t)$ and de-seasonalized local volatility $f(t)\sigma(t)$ where $\sigma(t)$ is an estimator of local volatility and $f(t)$ an estimator of the intraday periodicity (the so-called ``U-shape''). Throughout this paper, our statistical analyses will focus on $x(t)$, or on its ``jump-aligned'' version $\nosx(t) := x(t) \text{sign}(x(0))$, where $x(0)$ is the return corresponding to the jump. In other words,  $\overline{x}(t)$ is the rescaled return profile in the direction of the jump.

Under the null hypothesis of Gaussian residuals (no jump hypothesis) $|x(t)|$ converges towards a Gumbel distribution. A statistical test then allows us to reject the null hypothesis.
The resulting method comes down to detecting price movements where the z-score deviates by more than 4-sigma from their average value (here equal to zero).

The jump detection is performed on time-series describing individual stocks dynamics but also on averaged time-series across stocks belonging to the same sector. Hence, we obtain price jumps of individual stocks but also sectoral price jumps. 

Similarly to Marcaccioli \textit{et al.}~\cite{marcaccioli2022exogenous}, we find that price jumps are clustered in time. We assume that jumps taking place within the same ``time-cluster'' subsequent to an initial jump are merely replicas of the initial jump. They are likely to be either of the same dynamics (as they occurred for the same reason) or endogenously induced by the first jump of the cluster. We thus discard all the jumps that follows an initial jump. This leads to the same detection method as 
in~\cite{marcaccioli2022exogenous} which allows to retrieve an exponential distribution for the inter time between two consecutive initial jumps (see part II.D of \cite{marcaccioli2022exogenous}).

From such a collection of price jumps, we can then extract ``co-jumps''. A co-jump is simply defined as a set of jumps occurring in the same minute. Here we avoid tackling the question of lagged jumps and consider only simultaneous jumps (up to the minute resolution).

The price behavior before and after a jump can be used to classify the jump. 
In light of Marcaccioli \textit{et al.}'s findings (\cite{marcaccioli2022exogenous}), which indicate that volatility can begin to rise up to 75 minutes prior to the jump, we adopt a time window of 2 hours centered around the jump occurrence at time $t=\centr$. 
Consequently, for each jump we extract a time-series of 119 rescaled returns $x(t)$, corresponding to 1 hour preceding the jump and 1 hour following the jump.

We implement such detection on 301 US stocks from January 2015 to December 2022, considering only what happened between 10:30 and 15:00 in order to avoid special jumps due to the high activity at the beginning (due to people reacting to the overnight news/movement) and at the end of the day (due to market closing). In order to discard major market shocks, we also remove all co-jumps involving more than 250 stocks, and days on which the FED made an announcement (1 per month\footnote{see  \href{https://www.federalreserve.gov/newsevents/calendar.htm}{FOMC Calendars} }). We end up with 37 452 jumps, of which 16 127 belong to one of the 2534 co-jumps, and the remainder (21 325) are single jumps.

\subsection{Classification based on news labels}
\label{subsec:classif-news-labels}

In an attempt to characterize the reflexivity of a jump, one can gather the date and time of news associated to each stocks we consider\footnote{source: Bloomberg} and of the main US announcements\footnote{source: \href{https://www.investing.com/economic-calendar/}{economic-calendar}}. 
According to such news labels, we might label as ``news-related'' a jump which happened within 3 minutes of a news and label as ``non-news-related'' any other jump.
That would lead to a puny $\approx 4.3\%$ of the jumps being classified as ``news-related'' 
and is illustrated in Fig. \ref{fig:cojump_labels}. Hence, as previously argued in \cite{vol_news_jp,bormetti2015modelling,marcaccioli2022exogenous}, it appears that individual price jumps and more surprisingly co-jumps are often {\it not} related to news announcements. 

However, it is clear that some news may affect a whole economic sector and lead to a co-jump without appearing in our considered set of news. An example would be an OPEC announcement that affects oil prices and in turn ricochets onto stocks prices, without any of them explicitly showing up in the news feed. Another vivid example is the impeachment of the US president D. Trump in September 2019\footnote{For example, the largest co-jump is related to Nancy Pelosi announcement of a formal impeachment inquiry into US President Donald Trump. On 2019-09-24, at 14:13, 248 stocks saw their price jump in the same minute.}. Our ``news-related'' label is blind to such events. One objective of our study will be to propose a possible classification of co-jumps that does not rely on the news feed, see section \ref{sec:cojumps}.

%
\begin{figure*}
\centering
\includegraphics[width=0.9\linewidth]{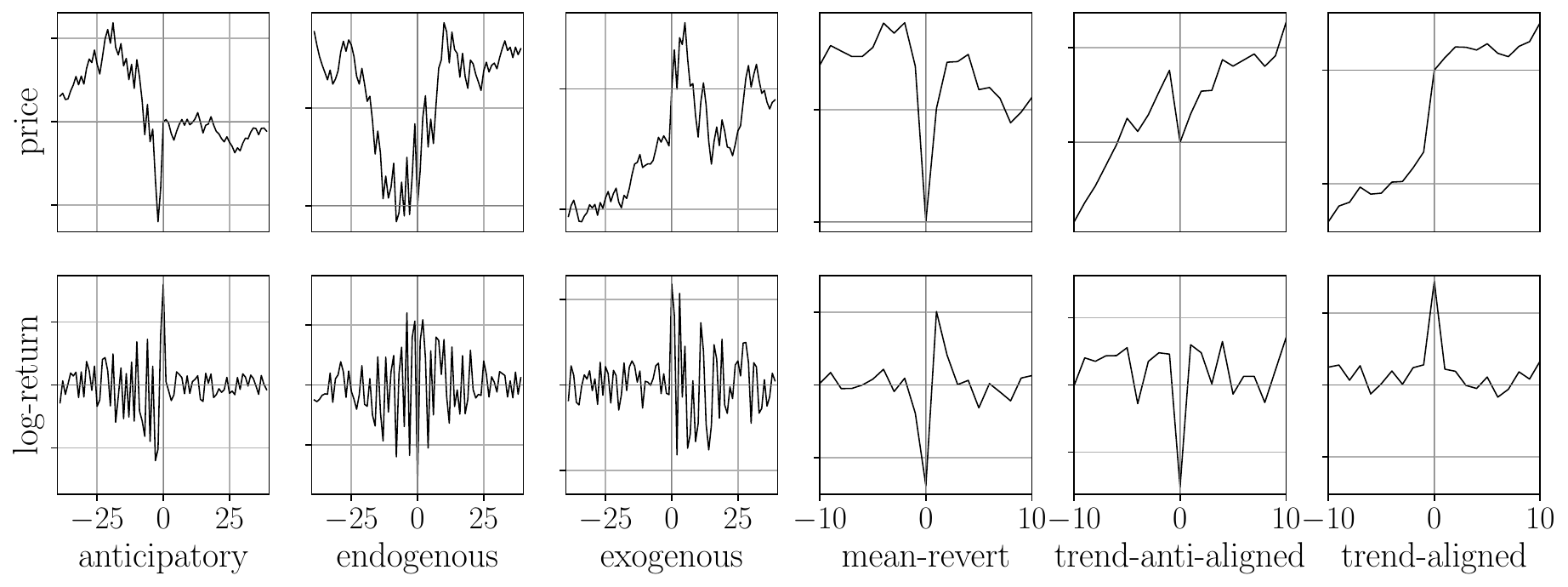}
\caption{
Classes of price jumps (synthetic examples). Each column shows an example of a class of jumps (price and log-return time-series).
The three first classes (anticipatory, endogenous, exogenous) are separated by measuring volatility asymmetry. The three last classes (mean-reverting, trend-anti-aligned, trend-aligned) are identified by analyzing the signed returns around the jump.
See Fig.~\ref{fig:examples-observed} for examples of true observed jumps.
}
\label{fig:examples}
\end{figure*}

\subsection{Classification based on the volatility profile}
\label{subsec:classification_volat}

In~\cite{marcaccioli2022exogenous}, Marcaccioli \textit{et al.} built a supervised classification of univariate jumps into exogenous and endogenous classes. The classification relies on parameters derived from fitting $|x(t)|$ to the following functional form \cite{deschatres2005dynamics}: 
\begin{equation}\label{eq:synthetic_jump_benchmarck}
    |x(t)| = \mathbf{1}_{t<t_c}\frac{N_<}{|t-t_c|^{p_<}} + \mathbf{1}_{t>t_c}\frac{N_>}{|t-t_c|^{p_>}} + d
\end{equation}
and on a measure of the asymmetry of the jump, defined as:
\begin{equation}\label{eq:asymmetry_def}
    \mathcal{A}_{\text{jump}} = \frac{\mathcal{A}_{>} - \mathcal{A}_{<}}{\mathcal{A}_{>}+\mathcal{A}_{<}}
\end{equation}
where $\mathcal{A}_{</>} := \sum_{t<\centr/t>\centr} |x(t)-\min_{t<\centr/t>\centr}(x(t))|$. Such an indicator means that when the activity is stronger before (resp. after) the jump, one has $\mathcal{A}_{\text{jump}}<0$ (resp. $\mathcal{A}_{\text{jump}}>0$). 
The classification is then obtained as a logistic regression of the news label (endogenous/exogenous) by the parameters $(\mathcal{A}_{\text{jump}},p_<,p_>,N_<,N_>,t_c)$.
Exogenous jumps appear as strongly asymmetric jumps with little activity ahead of the jump, i.e. 
$\mathcal{A}_{\text{jump}}>0$, whereas self-exciting endogenous jumps are much more symmetric with $\mathcal{A}_{\text{jump}} \approx 0$ \cite{marcaccioli2022exogenous}.

The above approach, based on news labels, presents several limitations:
\begin{itemize}
\item The classification partly relies on the goodness of fit of a power-law function (\ref{eq:synthetic_jump_benchmarck}), which is not assured. As a consequence, Marccacioli et al.~\cite{marcaccioli2022exogenous} restrict their study to only $\sim$ 5000 jumps out of the $\sim$ 37000 in the dataset, for which such a fit is acceptable. 
\item As discussed above, news labels might miss some  relevant economic news, so the resulting price jumps might be wrongly labeled as ``non news-related''. 
\item Exogenous jumps could have two types of dynamics: if the exogenous shock is a complete surprise, there should indeed be no activity before the jump. However, if the announcement is planned or if there was some news leakage, there might be a growth of activity before the jump. In this case, one would wrongly classify a news-related jump  as endogenous based on its approximately symmetric activity profile. 
\end{itemize}
In light of such limitations and in order to uncover new classes of jumps, beyond the sole study of their reflexivity, we opt in the rest of the paper for an unsupervised classification which significantly improves upon the method of \cite{marcaccioli2022exogenous} while still leaving open some ambiguities, as we will see below. 

In the following, although news labels do not reveal the whole truth about the reflexive nature of a jump, we will still call ``news-related'' jumps that occurred within 3 minutes of a news present in our database and ``non news-related'' all the others.

\section{\label{sec:SecUnivariateJumpClassif} 
Classification of single jumps using wavelets
}


The rescaled return time-series around a jump $x(t)\in\R^T$ is inherently noisy.
Relevant features $\Phi(x)\in\R^q$ must be extracted to effectively distinguish different classes of jumps.
Such features should be selected carefully, in particular, they should include time-asymmetry measures. Indeed, authors in~\cite{sornette2003endogenous,sornette2004endogenous,deschatres2005dynamics,crane2008robust,marcaccioli2022exogenous} show that the jumps mostly differ in their time-asymmetry: endogenous jumps tend to be more symmetric around the jump than exogenous ones. But what are the other possibly relevant features?
In this section, we embrace a signal processing approach to discover important features of univariate jumps and unveil new classes of jumps that are prevalent in the data.

\subsection{Wavelet and scattering coefficients}
\label{subsec:wavelet-representation}

\begin{figure}[h!]
\centering
\includegraphics[width=0.9\linewidth]{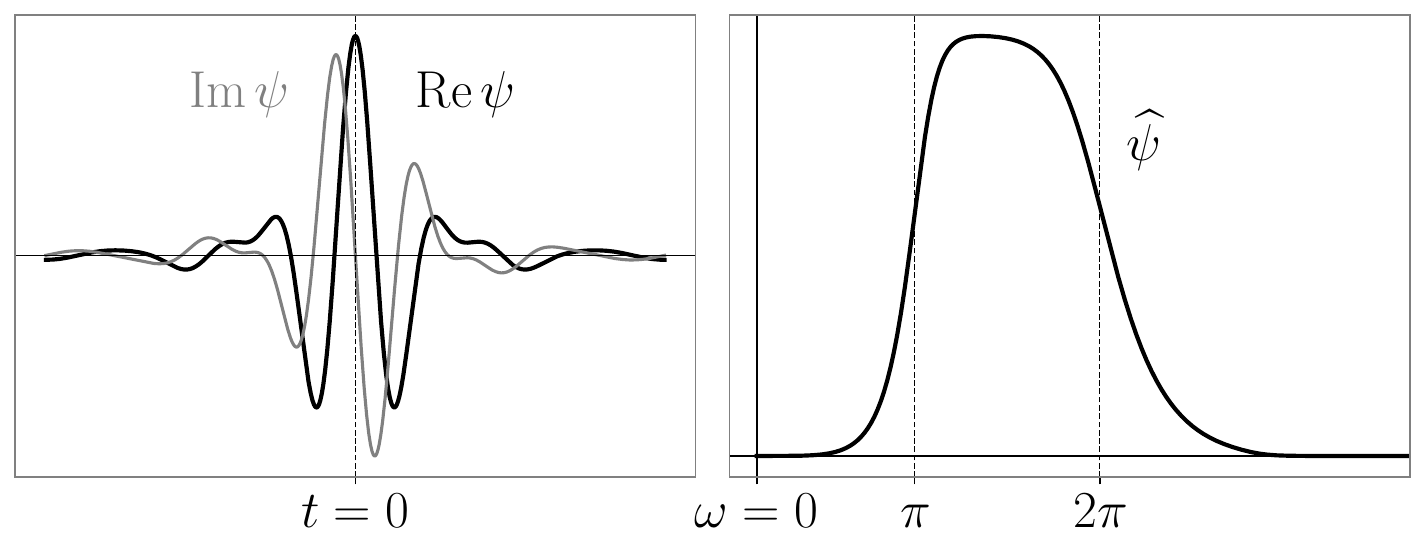}
\caption{Filter used to analyze jump time-series.
Left: complex Battle-Lemarié wavelet $\psi(t)$ as a function of $t$.
Right: Fourier transform $\widehat{\psi}(\om)$ as a function of $\om$.}
\label{fig:wavelet}
\end{figure}

Wavelet filters have been used to analyze and classify transient events, see e.g.~\cite{probert2002detection,kim2015wavelet,rueda2018transient,cuoco2018wavelet}.
A complex wavelet filter $\psi(t)$ is a filter whose Fourier transform 
$\widehat{\psi}(\om) = \int \psi(t)\, e^{-i \om t}\, {\rm d}t$, is real. 
It is localized both in time and Fourier domains, see Fig.~\ref{fig:wavelet}.
It has a fast decay away from $t=0$ and a zero-average $\int \psi(t)\,{\rm d}t = 0$.
We write $\psi(t)={\rm Re}\,\psi(t)+i\,{\rm Im}\,\psi(t)$ where ${\rm Re}\,\psi(t)$ and ${\rm Im}\,\psi(t)$ are its real and imaginary parts. They are respectively even and odd functions:
\begin{equation}
\label{eq:re-im-wavelet}
{\rm Re}\,\psi(-t) = {\rm Re}\,\psi(t) ~~\text{and}~~ 
{\rm Im}\,\psi(-t) = -{\rm Im}\,\psi(t).
\end{equation}
The wavelet coefficients $W_jx(t)$ compute the variations of the signal $x$ around $t$ at scale $2^j$, for $j=1,\ldots, J$ with
\begin{equation}
\label{eq:WX}
W_jx(t) := x\star\psi_j(t)~~\mbox{where}~~\psi_j(t) = \psi(2^{-j}t).
\end{equation}
where $\star$ denotes the convolution: $x\star y(t) := \int x(t-u)y(-u)\,{\rm d}u$.

The sign of the jump $\text{sign}(x(0))$ and its amplitude $|x(0)|$ vary, but they are not necessarily informative for their classification. To remove this source of variability we consider the {\it jump-aligned} time-series 
\begin{equation}
    \nosx(t) = \text{sign}(x(0))\,x(t)
\end{equation}
and we further normalize the wavelet coefficients (\ref{eq:WX}) by the corresponding ``volatility'' $\sigma_j$ of the full time-series, defined as $\sigma_j^2 = \Av{t}{|x\star\psi_j(t)|^2}$, where $\Av{t}{\cdot}$ denotes the empirical average over time $t$.

From Eq.~(\ref{eq:re-im-wavelet}), one can see that if $x$ is an even signal i.e. $x(-t)=x(t)$ then $\text{Im}\,Wx(t,j) \equiv 0$. This property is key to detect asymmetry of a signal at different scales. 

Volatility information can be extracted by taking a modulus. 
The time-series $|W_j x(t)|$ provides the volatility of the signal at scale $2^j$. 
This volatility can be asymmetrical in $t=0$. In order to quantify it, we again consider the wavelet coefficients at $t=0$
\begin{equation}
\label{eq:WmWx}
W_{j_2}|W_{j_1}x|(t) := |x\star\psi_{j_1}|\star\psi_{j_2}(t).
\end{equation}
%

Our representation for univariate jumps in this paper is thus composed of wavelet coefficients (\ref{eq:WX}) at $t=\centr$ and scattering coefficients (\ref{eq:WmWx}) at $t=\centr$
\begin{equation}
\label{eq:scat-coeffs}
\Phi(x) = \bigg(W_j\nosx(\centr)\,,\, W_{j_2}|W_{j_1}x|(\centr)\bigg).
\end{equation}
For a time-series of size $T$, it contains less than $(\log_2 T)^2/2$ coefficients which represents few coefficients. 
In our case, $T=119$ and we chose $J=6$, which yields 42 coefficients (21 real parts and 21 imaginary parts). 
The normalized scattering features $\Phi(x)$ (Eq.~\eqref{eq:scat-coeffs}) are invariant to sign change and to dilation
\begin{equation}
\nonumber
\Phi(-x) = \Phi(x) ~~ \text{and} ~~ \Phi(\lambda x) = \Phi(x).
\end{equation}
which means we do not aim at discriminating jumps neither based on their sign nor on their amplitude. 

In order to classify price jumps, we are interested in Principal Component directions of the 42-dimensional vector $\Phi(x)$ in the dataset. This method, called kernel PCA~\cite{scholkopf1997kernel}, relies on the linear separation power of our scattering coefficients $\Phi(x)$.
We considered several directions, i.e. combinations of scattering coefficients, and found three salient features: the time-asymmetry of the volatility, the mean-reversion and the trend behavior of the price around the jump.

\subsection{First Direction $D_1$: Volatility asymmetry
}\label{subsec:reflexivity}

\subsubsection{Three types of jumps}

\begin{figure}[h]
\centering
\includegraphics[width=\linewidth]{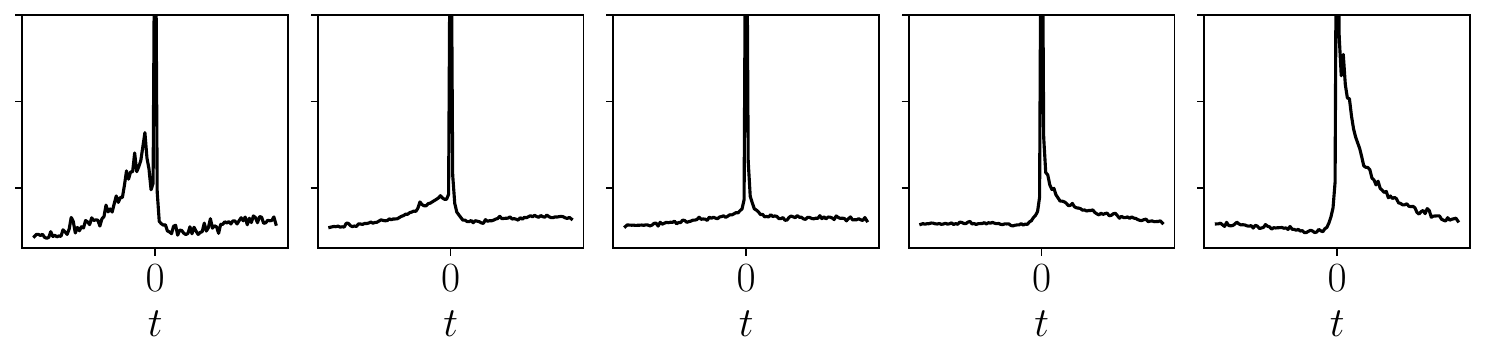}
\caption{
Average absolute profiles $|x(t)|$ of jumps along direction $D_1$ (sliced into five bins, delimited by quantiles 0.1, 0.25, 0.35, 0.9). From left to right: anticipatory jumps, endogenous jumps and exogenous jumps.}
\label{fig:proj_on_x_profiles}
\end{figure}

The first PCA direction (called $D_1$ henceforth) is a linear combination of the 15 coefficients $\text{Im}\,W_{j_2}|W_{j_1}x|(\centr)$ in Eq.~\eqref{eq:scat-coeffs}, which characterizes time-asymmetry of the {\it volatility profile} at multiple scales $2^{j_2}$, confirming previous analysis that postulated this asymmetry to be relevant. Such a linear combination  allows one to embed each jump time-series into a one dimensional space, which quantifies the reflexive nature of each jump. In fact, Fig.~\ref{fig:proj_on_x_profiles}
and Appendix \ref{app:abs_grid} 
display average profiles $|x(t)|$ along the ``reflexive direction'' $D_1$. One can visually verify that such a representation discriminates jumps according to the asymmetry of their profiles as measured by $\mathcal{A}_{\text{jump}}$ (Eq.~\eqref{eq:asymmetry_def}): the $D_1$ direction continuously separates asymmetric jumps with dominant activity before the shock from asymmetric jumps with dominant activity after the shock; see Figs. \ref{fig:proj_on_x_profiles}, \ref{fig:abs_grid} and \ref{fig:projection_asym}.

From this analysis, three types of jumps can thus be defined:
\begin{itemize}
    \item Asymmetric jumps with dominant activity {\it before} the shock. This type of jumps, which we call ``anticipatory'', was quite unexpected and was not discussed in \cite{marcaccioli2022exogenous}. 
    \item Symmetric jumps, with an pre-shock excitation activity that approximately mirrors the post-shock relaxation activity. These were called ``endogenous jumps'' in \cite{marcaccioli2022exogenous}: increased activity before the jump is in fact responsible for the jump itself, with some decay of activity thereafter. The symmetry of the profile for endogenous jumps is in fact predicted by a Hawkes process description of the self-exciting mechanism, see \cite{deschatres2005dynamics, marcaccioli2022exogenous}.
    \item Asymmetric jumps with dominant activity after the shock. These were called ``exogenous jumps'' in \cite{marcaccioli2022exogenous}: the market reacts  {\it after} unexpected news, but not before.
\end{itemize}
%
%
%

%
In order to validate the above analysis, we created synthetic time-series with volatility profiles of varying time-asymmetry and applied our classification method. Results of this benchmark case are shown in Appendix \ref{app:benchmark}, and fully confirm that the $D_1$ direction indeed separates jumps according to their asymmetry $\mathcal{A}_{\text{jump}}$.  



\subsubsection{Discussion}
Using the above classification, we find that a large proportion ($\sim 50\%$) of our sample exhibit positive asymmetry and should naively be considered as exogenous jumps. This seems in contradiction with the results of \cite{marcaccioli2022exogenous}, where exogenous jumps were found to be a minority, and with a fraction of jumps associated to a news found to be $4.3 \%$, as already quoted above. Several arguments can explain such a difference. 
\begin{itemize}
     \item 
The main one is the fact that our analysis includes all jumps involved in a sector jump (corresponding to $24 \%$ of all jumps) whereas those jumps were discarded in \cite{marcaccioli2022exogenous}. Sector jumps are such that many stocks of the same industry jump simultaneously. While some of these jumps are likely due to major exogenous shocks -- like macro-economic announcements -- that affect a whole economic sector or even the whole market, we argue in section \ref{sec:cojumps} that these jumps can actually be induced by a jump of one particular stock of the sector, which is deemed as ``news'' in and by itself. In any case, taking these sector jumps into account mechanically increases the count of jumps with a positive $D_1$ score. In the present study, we chose to keep these co-jumps and study their statistics, to which we will specifically turn in section \ref{sec:cojumps}. 
\item As already noted above,  the classification of single jump profiles in~\cite{marcaccioli2022exogenous} relies on the goodness of fit of power law function \eqref{eq:synthetic_jump_benchmarck}, and as such, was only conducted on a smaller sample for which such a fit is acceptable ($\sim 5000$ jumps out of $\sim 37 000$ jumps).

\end{itemize}
The appearance of ``anticipatory jumps'', where the asymmetry parameter $\mathcal{A}_{\text{jump}}$ (see Eq.~\eqref{eq:asymmetry_def}) is negative, came somewhat as a surprise to us. One possible interpretation is that these jumps are in fact also endogenous, with a pre-shock self-exciting dynamics and very little ``after-shocks''. Indeed, if such jumps are immediately deemed endogenous by the market, it might make sense that activity quickly reverts back to normal. This would simply mean that the Hawkes framework predicting a symmetric profile is not adapted to describe all endogenous shocks. 

Another possibility is that such events correspond to news/exogenous events whose {\it timing} is expected by the market, which leads to increased activity before the actual release time. But if the actual news content turns out to be insignificant, it would again make sense that the market activity quickly wanes off. We in fact find a very small fraction of news-related jumps with $D_1 < 0$, see in Fig. \ref{fig:proj_qu}, bottom graph.

\subsection{Second Direction $D_2$: Mean-Reversion}

\subsubsection{Capturing Mean-Reversion}

We observed that coefficients $\text{Im}\,W_{j_1}\nosx(0)$ (\ref{eq:scat-coeffs}) for fine scales, i.e. small $j_1$, are consistently chosen by the leading PCA directions. They amount to multiplying the jump-aligned time-series $\nosx(t)$ by the imaginary filter $\text{Im}\,\psi_1(t)$ (see Fig.~\ref{fig:wavelet}) and averaging over $t$. 
Such coefficients capture the asymmetry of the return profile shortly before and shortly after the jump, and define what we will call below direction $D_2$.

A typical time-series that maximizes this coefficient is thus characterized by a positive value of $\nosx(-1)$ and a negative value of  $\nosx(1)$. In other words, large positive values along the $D_2$ coordinate capture mean-reverting return profiles, i.e. positive (resp. negative) returns before a positive (resp. negative) jump that become negative (resp. positive) immediately after the jump. 

Large negative values along the $D_2$ coordinate, on the other hand, also capture mean-reverting return profiles, but in this case mean-reversion starts with (or is triggered by?) the jump itself, and not after the jump.  
\begin{figure}
\centering
\begin{subfigure}[t]{0.48\linewidth}
    \centering
    \includegraphics[width=\linewidth]{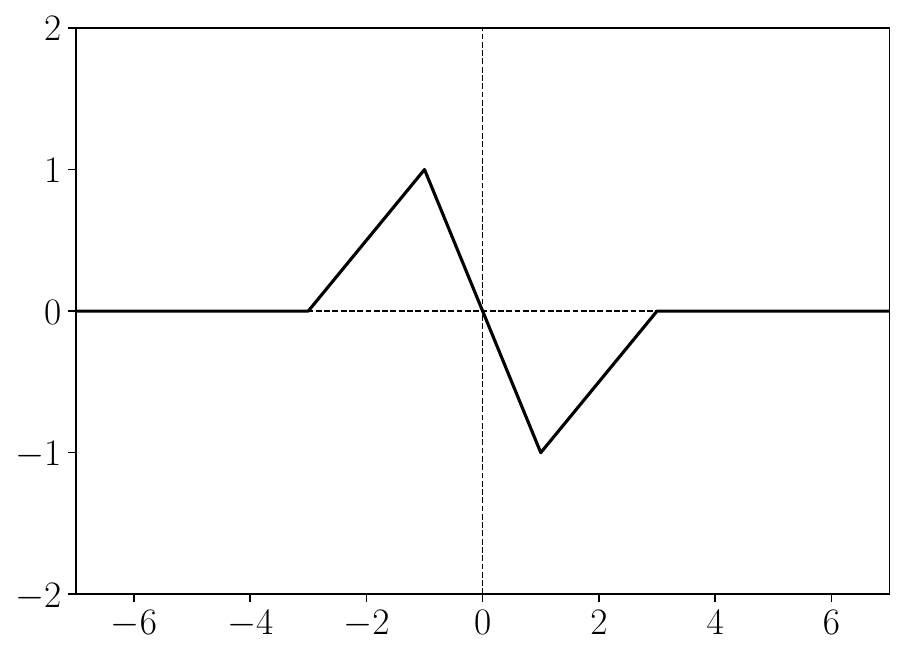}
    \caption{$\psi_\text{MR}$}
\end{subfigure}
\begin{subfigure}[t]{0.48\linewidth}
    \centering
    \includegraphics[width=\linewidth]{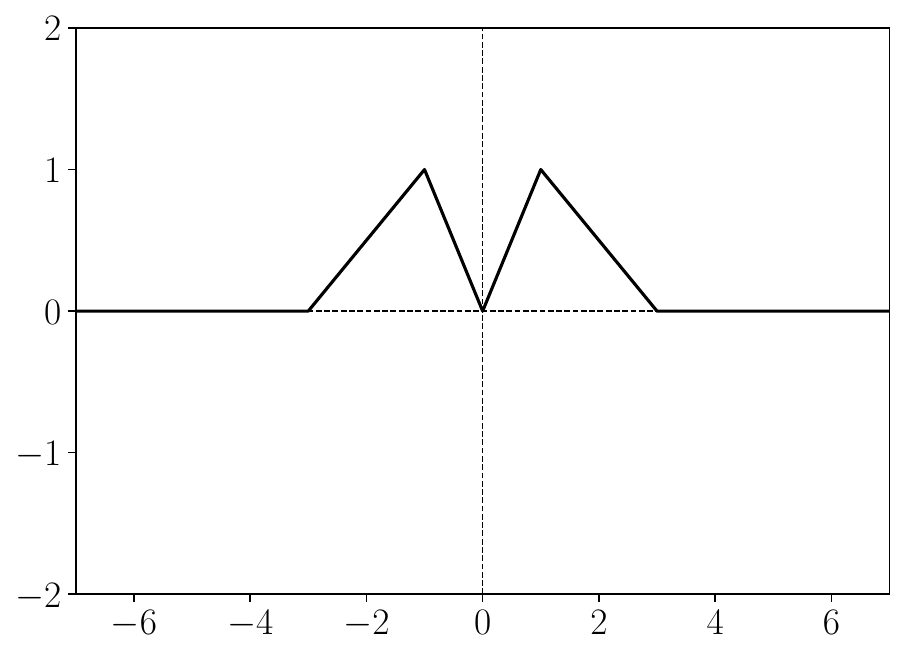}
    \caption{$\psi_\text{TR}$}
\end{subfigure}
\caption{
Handcrafted filters for measuring the mean-reversion (filter $\psi_\text{MR}$) or the trend (filter $\psi_\text{TR}$) character of a jump. 
Average profiles along resulting mean-reversion and trend directions are shown 
in Fig.~\ref{fig:mr-projection} and Fig.~\ref{fig:td-projection}.
}
\label{fig:handcrafted-filter}
\end{figure}

Now that we identified a potentially discriminating direction using PCA, we transition to a simpler filter tailored to capture short time mean-reversion, depicted in Fig.~\ref{fig:handcrafted-filter}. This filter is then applied to the jump-aligned time-series $\nosx(t)$
\begin{equation}
\label{eq:mr}
\widetilde{D}_2(x) := \nosx\star\psi_\text{MR}(0),
\end{equation}
where the tilde indicates that we have simplified the true second PCA direction and only retained the component spanned by $\psi_\text{MR}$.

\subsubsection{A 2D representation of jumps}

\begin{figure}[h!]
\centering
\includegraphics[width = \linewidth]{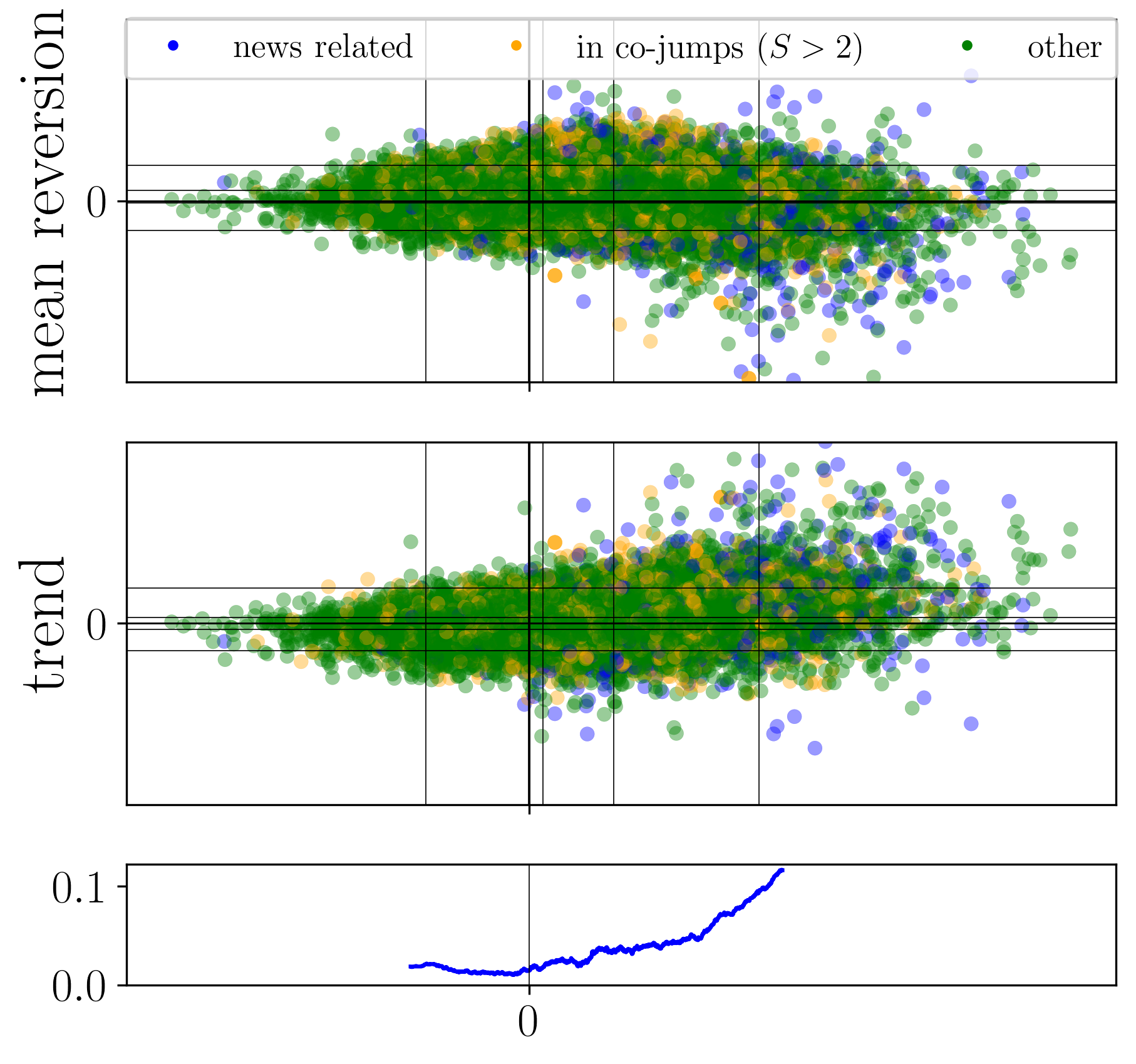}
\caption{
\underline{Top graph}: Projection of jumps in our dataset onto the reflexive direction $D_1$ (horizontal axis) and mean-reverting direction $\widetilde{D}_2$
(vertical axis)
\underline{Middle graph}: Projection of our dataset on the reflexive direction $D_1$ (horizontal axis) and trend direction $\widetilde{D}_3$
(vertical axis).
Each point represents a jump, the blue color corresponds to news-related jumps according  to the classification of Section \ref{subsec:classif-news-labels}, the oranges are jumps involved in a co-jump of size greater than 2 and non news related and the greens are all the other jumps.
The vertical and horizontal lines represent the following quantiles: 0.05, 0.35, 0.65, 0.95. 
\underline{Bottom graph}: ratio of ``news-related'' jumps along the reflexive direction $D_1$, based on a direct classification using the news feed (rolling ratio every 2000 jumps).
}
\label{fig:proj_qu}
\end{figure}

Based on the first volatility asymmetry direction $D_1$ and the mean-reversion direction $\widetilde{D}_2$, we are in a position to propose the 2D representation of jumps shown in Fig.~\ref{fig:proj_qu} (top), in which the horizontal axis corresponds to $D_1$ and the vertical axis corresponds to the mean-reversion index $\widetilde{D}_2$. 
Visually, news-related jumps are mostly to the right of the projection, corresponding to increased volatility after the jump, as expected.

\begin{figure}[H]
\centering
\includegraphics[width = \linewidth]{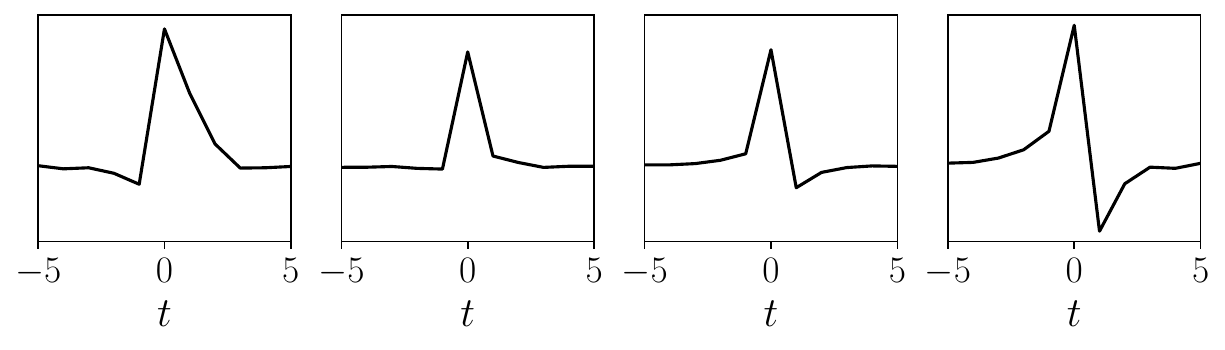}
\caption{
\underline{Mean-reverting profiles}. Average jump-aligned return profiles $\nosx(t)=\text{sign}(x(\centr))\,x(t)$ along the mean-reverting direction $\widetilde{D}_2$ (sliced into four bins, delimited by quantiles 0.1, 0.5 \& 0.9). Left-most graph: price jumps mean-revert on previous trends. Right-most graph: prices mean-revert after the jump. 
}
\label{fig:mr-projection}
\end{figure}

In order to illustrate the discriminating power of such coefficient, Fig.~\ref{fig:mr-projection} displays the average profiles of $\nosx(t)$ along the $\widetilde{D}_2$ axis. One can see that jumps with a high coefficient $\widetilde{D}_2$ (rightmost graph) are characterized by a strong pre-jump trend aligned with the jump, followed by a change of sign in the next minute {\it after} the jump (as also shown in Appendix \ref{app:signed_grid_mr}). 

The leftmost graph, on the other hand, shows relatively mild pre-jump trends opposite to the jump, followed by stronger trends in the direction of the jump, not very different from the cases corresponding to quantiles between 0.1 and 0.5. 
In our dataset, 60\% of the jumps have a positive mean-reversion score $D_2>0$; we refer to Fig.~\ref{fig:imagW_dis} in Appendix \ref{app:dis_j1} for the full distribution of $D_2$. 

To confirm this observation and ascertain that it is not attributable to spurious effects in the data processing, we looked deeper into these jumps. To get a better understanding of the mechanisms at play, we investigated what happens at tick-by-tick scale in the Limit Order Book. We show in Appendix \ref{app:LOBexample} two illustrative examples, see Fig. \ref{fig:LOBillustration}. We again observe, at a different time resolution, a strong mean-reversion behavior induced by order placement. 
Note that both exogenous, or endogenous jumps can have such mean reverting behavior, as clear from the 2D representation Fig. \ref{fig:proj_qu}.

In fact, a mean reverting behavior can be expected both following an exaggerated response to a news release, or after a self-initiated jump with no discernible catalyst.
This is confirmed by Fig.~\ref{fig:corr-refl-mr-td} which shows positive average values of $\widetilde{D}_2$ for all levels of reflexivity $D_1$, except for strongly exogenous jumps (large values of $D_1>0$), where the mean-reversion disappears ($\widetilde{D_2}\approx0$). 

Note finally that mean-reversion is characterized by a V-shape price profile (see Fig. \ref{fig:LOBillustration}), which has recently been used as a criterion to detect price jumps in time-series (\cite{flora2022v}).

\subsection{Third Direction $D_3$: Trend}

In the previous section, we have defined a filter $\psi_\text{MR}$ that detects mean-reversion, but is by construction orthogonal to trends, i.e. post-jump returns continuing in the same direction as pre-jump returns. This feature can be naturally captured by the trend filter $\psi_\text{TR}$ shown in Fig.~\ref{fig:handcrafted-filter}, which is orthogonal to the mean-reversion filter $\psi_\text{MR}$. This filter is then applied to the jump-aligned profile $\nosx(t)$ to get the following trend score
\begin{equation}
\label{eq:trend}
\widetilde{D}_3(x) := \nosx\star\psi_\text{TR}(0).
\end{equation}
A large positive value of $\widetilde{D}_3(x)$ therefore describes a persistent trend aligned with the direction of the jump. If such jumps exist, we refer to them as ``trend-aligned" jumps. A large negative value of $\widetilde{D}_3(x)$ indicates that the jump goes against the pre- and post-jump trend. If such jumps exist, we refer to them as ``trend-anti-aligned" jumps. 

\begin{figure}[H]
\centering
\includegraphics[width=\linewidth]{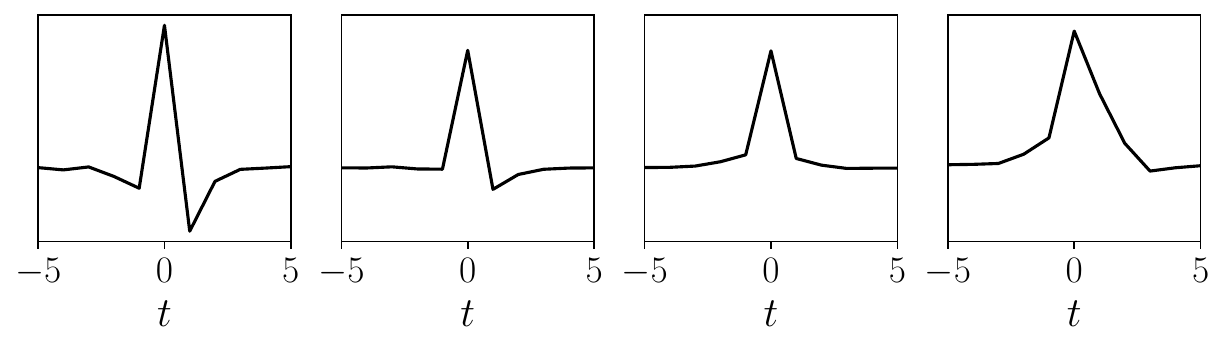}
\caption{
\underline{Trending profiles}. Average jump-aligned return profiles $\nosx(t)=\text{sign}(x(\centr))\,x(t)$ along the trend direction $\widetilde{D}_3$ (again sliced into four bins, delimited by quantiles 0.1, 0.5 \& 0.9). Left-most graph: anti-aligned trends. Right-most graph: aligned trends.
}
\label{fig:td-projection}
\end{figure}

Fig.~\ref{fig:td-projection} shows that both classes of jumps do indeed exist: the average profiles in the first and last quantiles in Fig.~\ref{fig:td-projection} do conform to expectations. Furthermore we directly observe many stylized examples such as the one reported in Fig.~\ref{fig:examples-observed}. As for the mean-reversion indicator, we can represent all jumps in 2D plane based on $D_1$ and $\widetilde{D}_3$ (see the bottom graph in Fig.~\ref{fig:proj_qu}). Visually, trending news-related jumps appear to be mostly aligned with the jump (top-right corner), although anti-aligned trends can also be spotted for moderate values of $D_1$. Different profiles of $\nosx(t)$ corresponding to the grid are shown in Appendix \ref{app:signed_grid_tr}.  


%
%
\subsection{Preliminary Conclusions}

Let us summarize the results obtained by our unsupervised approach so far. First, our proposed 2D projections provide an embedding of a jump according to three meaningful, intuitive properties: its self-reflexive nature (along horizontal axis), its mean-reversion character or its trend character (along vertical axis). On top of the separation between exogenous and endogenous jumps, our clustering method revealed new classes of jumps, some of which we did not expect a priori:  anticipatory jumps, mean-reverting jumps, trend-aligned and trend-anti-aligned jumps. 
Identifying additional interpretable classes of jumps might be possible by considering more expressive wavelet-based embeddings such as Scattering Spectra recently used in the context of financial time series~\cite{morel2022scale,morel2023path}.
However, our attempts so far seemed to mostly recover directions which overlap with the volatility time-asymmetry and mean-reverting directions.


\section{\label{sec:cojumps} Classification of co-jumps}

A ``co-jump'' is defined as a collection of jumps across several stocks, occurring in the same minute.
The number  $S$ of assets involved in the co-jump is referred to as the ``size'' of the co-jump. 
Co-jumps reveal inter-connectivity and contagion in financial markets \cite{gerig2012high, bormetti2015modelling, calcagnile2018collective}. As such, studying them -- in particular their possible reflexive nature -- is a crucial question for investors and regulators alike. 
This section aims at investigating whether co-jumps are created through endogenous dynamics or exogenous shocks. 

\begin{figure}
\centering
\begin{subfigure}[t]{0.48\linewidth}
    \centering
    \includegraphics[width=\linewidth]{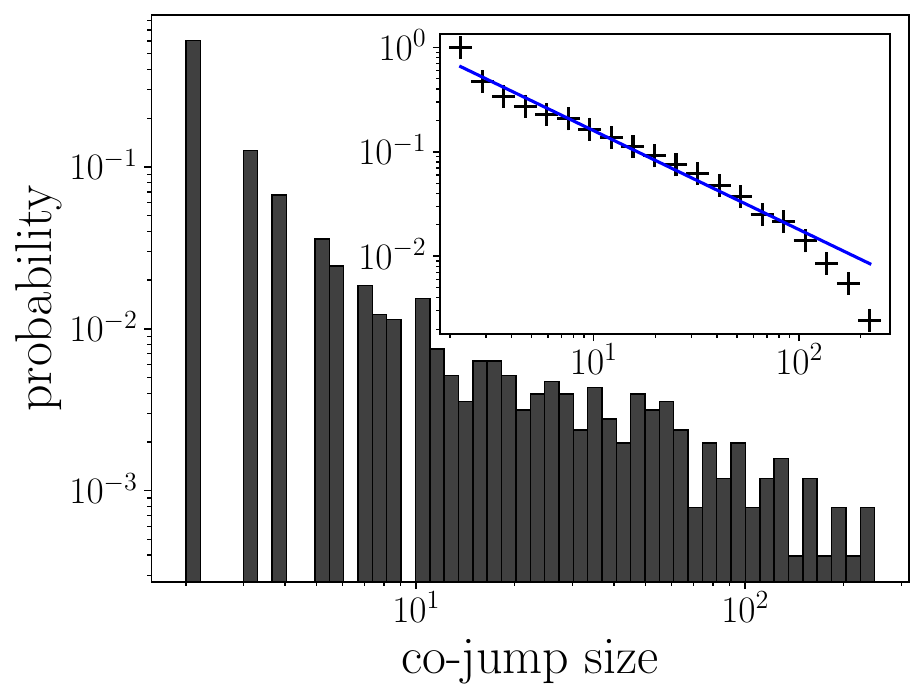}
    \caption{Co-jump size}
    \label{fig:size_dis}
\end{subfigure}
\begin{subfigure}[t]{0.48\linewidth}
    \centering
    \includegraphics[width=\linewidth]{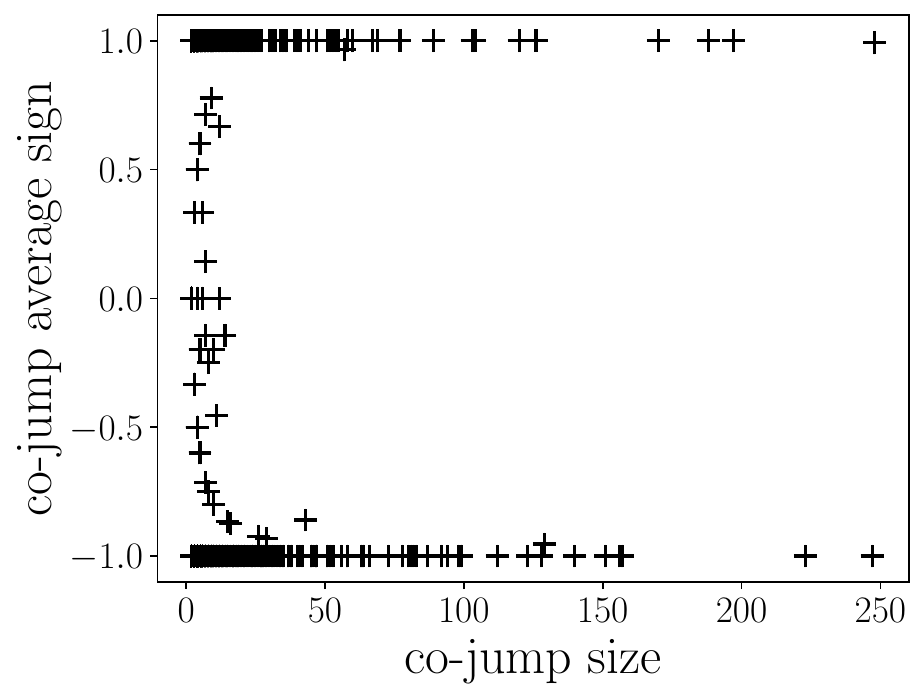}
    \caption{Co-jump sign}
    \label{fig:size_VS_signs}
\end{subfigure}
\caption{Statistics on co-jumps. (a) Main: Distribution of co-jumps size i.e. number of stocks involved in a co-jump. Inset: Cumulative distribution of co-jumps size for co-jumps with $\min(D_1)<0$ and $\min(D_1) < \overline{D_1}-1\sigma$, defining the LL and LR regions in Fig. \ref{fig:meanVSmin}.
The slope of the fit in log-log coordinate (plain line in blue) is $-\tau=-0.95$. Notice that the data bends down faster for large $S$.
(b) Average sign of jumps involved in a co-jump, showing that most co-jumps are composed of jumps in the same direction.}
\label{fig:co-jumps-stats_0}
\end{figure}

To assemble our co-jump dataset we consider the same dataset of jumps as in the previous section.
We end up with 2534 co-jumps, the size of which varies from 2 stocks to 248 stocks. The co-jumps cumulative size distribution, restricted to endogenous jumps, is shown in Fig. \ref{fig:size_dis}, inset. 
Quite remarkably, the tail of this distribution is well fitted by a power-law $S^{-\tau}$ with exponent $\tau \approx 1$, with a cut-off for $S \gtrsim 100$. As we discuss in Appendix \ref{app:branching}, such a value for $\tau$ can be rationalized within the framework of critical branching processes \cite{harris1963theory}, as if co-jumps were the result of a contagion mechanism. Such a power-law behaviour was already noted in previous works: in Ref. \cite{vol_news_jp} on a US data set from 2004 to 2006,  in \cite{calcagnile2018collective} from 2001 to 2013 and in \cite{aubrun2023multivariate} from 2013 to 2018. 

The signs of the jumps involved in a co-jump are, most of the time, all aligned, i.e. different stocks jump in the same direction, as shown in Fig.~\ref{fig:size_VS_signs}.


%
%

\begin{figure}[h!]
\centering
\includegraphics[width=\linewidth]{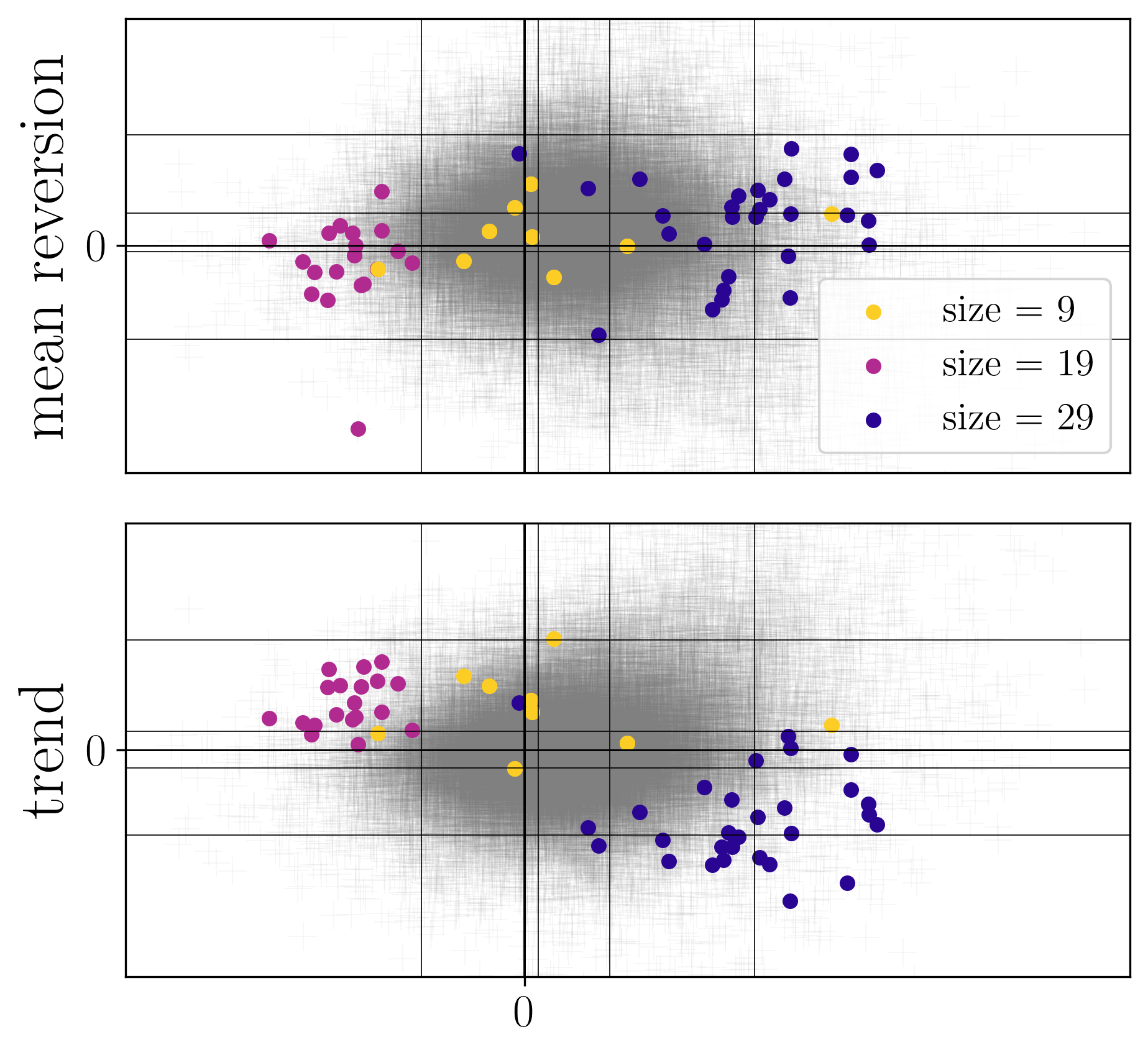}
\caption{
Projections of 3 co-jumps along our 2D projections.
\underline{Yellow co-jump}: one jump is exogenous and the others are more endogenous. \underline{Pink co-jump}: all jumps of the co-jumps are endogenous and are trend-aligned. \underline{Purple co-jump}: Most jumps appear to be exogenous except one. Those jumps are also trend-anti-aligned. 
}
\label{fig:co_jump_proj_ex}
\end{figure}

The first stage of co-jump characterization is to classify jumps according to their reflexivity coordinate along the $D_1$ direction. 
In Fig. \ref{fig:co_jump_proj_ex}, we highlight the coordinates of three particular co-jumps in the 2D projections introduced in the previous section. Each color point is a stock involved in one of the three co-jumps. Let us comment on each of these three cases in turn:
\begin{itemize}
\item The purple co-jump, with 29 stocks involved, has most of its elements in the right side of the 2D projection, suggesting an exogenous, news driven shock. However, one of the jump is below the 0.35 quantile and therefore appears endogenous. This might be a mis-classification because of the inherent noise in our $D_1$ reflexivity score. An alternative interpretation might however be that this particular stock jumped for no particular reason and this created a surprise  to which other stocks reacted. 
\item The pink co-jump, with 19 stocks involved, staunchly belongs to the anticipatory class -- which we believe to be of endogenous nature, as explained above. Co-jumps with a negative or positive but moderate maximum value of the $D_1$ score can thus be deemed endogenous.
\item The yellow co-jump, with 9 stocks involved, has most of its elements in the intermediate ``endogenous" region, except one which is classified as exogenous. This might be either again a mis-classification because of the inherent noise in our $D_1$ reflexivity score, or else a stock that was not part of the anomalous pre-jump activity but is drawn into the jump through contagion. 
\end{itemize}
 

\begin{figure}
\centering
\includegraphics[width=\linewidth]{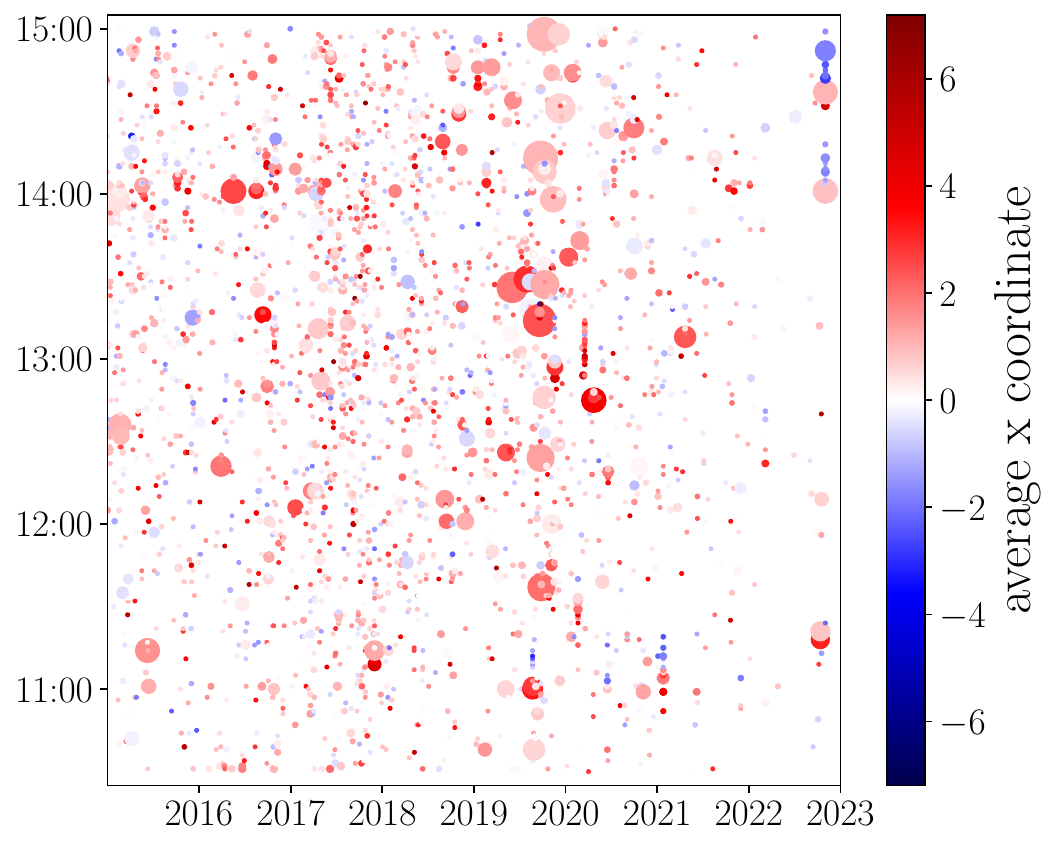}
\caption{
Reflexive score $\overline{D_1}$ of co-jumps in our dataset, obtained by averaging the reflexive score $D_1$ of each jump involved in a co-jump.
Large co-jumps tend to have a higher average score (in red) but, surprisingly, there many large co-jumps with pale color  that would be classified as endogenous. See discussion in the text.
}
\label{fig:cojump_x_av}
\end{figure}

From these cursory observations, one may propose three natural indicators for classifying co-jumps:
\begin{enumerate}
\item The average value of the individual reflexivity score $\overline{D_1}$ over all jumps belonging to a given co-jump, see Fig. \ref{fig:cojump_x_av}.
\item The maximum value of the individual reflexivity score $D_1$ over all jumps belonging to a given co-jump: if the most exogenous jump is still deemed endogenous, the whole co-jump is classified as endogenous (see distribution in Fig. \ref{fig:distribution_max_x}).
\item The minimum value of the individual reflexivity score $D_1$ over all jumps belonging to a given co-jump: if the most endogenous jump is still deemed exogenous, the whole co-jump is classified as exogenous (see distribution in Fig. \ref{fig:distribution_min_x}).
  

\end{enumerate}

\begin{figure}
    \centering
    \includegraphics[width=\linewidth]{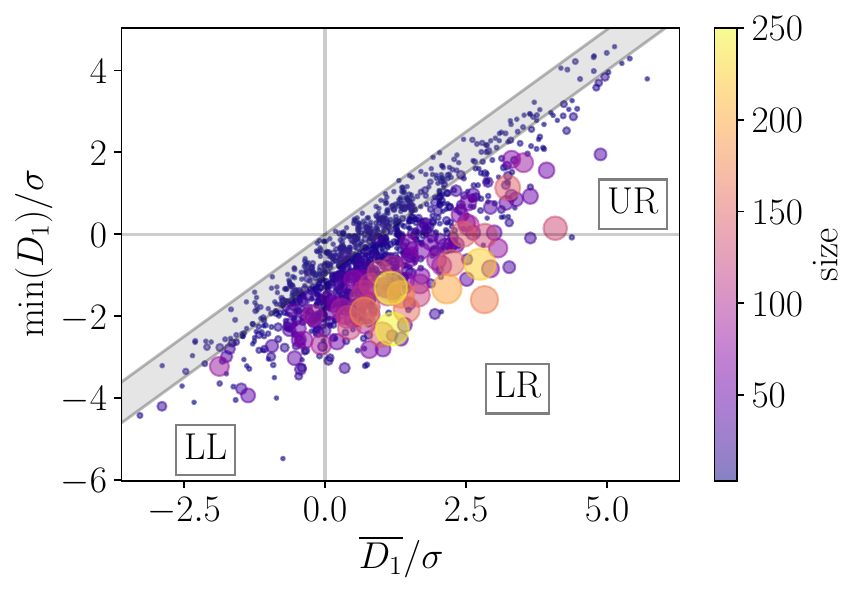}
    \caption{
    Minimum value of reflexivity score $D_1$ over all jumps of a given co-jump as a function of the average value $\overline{D_1}$ of reflexivity score $D_1$ over all jumps of a given co-jump (co-jump indicator 3 as a function of co-jump indicator 1). Both indicators have been normalized by the average standard deviation of the reflexivity score $D_1$ of co-jumps with the same size $\sigma$. The size and color of a point depict the size of the co-jump. The grey area represents the zone between $\min(D_1)=\overline{D_1}$ and $\min(D_1)=\overline{D_1}-1$, corresponding to co-jumps where the difference between the minimum and the average $D_1$ score is less than 1$\sigma$. Here, we only consider co-jumps with a size strictly greater than 2. LL, LR \& UR stand for lower left, lower right and upper right.
    }
    \label{fig:meanVSmin}
\end{figure}

Fig. \ref{fig:meanVSmin} represents the normalized minimum value of reflexivity score $D_1$ over all jumps of a given co-jump as a function of the normalized average value of reflexivity score $D_1$ over all jumps of a given co-jump (co-jump indicator 3 as a function of co-jump indicator 1). The normalization is such that Fig.~\ref{fig:meanVSmin} can be read in units of standard deviation of the reflexivity score $D_1$ for co-jumps of same size, i.e. $\sigma$ is the average of the standard deviation of the score $D_1$ over co-jumps with same size. The size and color of a point depict the size of the co-jump. The gray shaded region represents jumps with insignificant differences between the mean and the minimum value of the $D_1$ score.

Co-jumps with negative minimum and average values of reflexivity score $D_1$ (lower left quadrant of Fig.\ref{fig:meanVSmin}, LL) can be deemed endogenous, whereas co-jumps with positive minimum and average values of reflexivity score $D_1$ (upper right quadrant of Fig.\ref{fig:meanVSmin}, UR) can be deemed exogenous. 

The lower right quadrant (LR) represent more intriguing co-jumps. Indeed, according to their average score $D_1$ those co-jumps should naively  be classified as exogenous, however they contain at least one strongly endogenous co-jump. It might be that those endogenous jumps, whose pre-activity starts while most other stocks are still quiet, are interpreted in and by themselves as news. This surprise triggers all other jumps -- which therefore appear as exogenous, with no special pre-jump activity but without being related to any news!

Note that the largest co-jumps are in the LR region; our interpretation in terms of a contagion mechanism would then naturally explain the power-law distribution of size $S^{-\tau}$ shown in Fig. \ref{fig:size_dis}. 



There are obviously also large sector wide co-jumps that are truly news-related -- upper-right quadrant of Fig. \ref{fig:meanVSmin}. For instance, the significant co-jumps highlighting the year 2019 mostly exhibit a negative average (exogenous) and are related to the announcements during the US vs China trade war. 

Conversely, some co-jumps (20\% of our sample) involve only jumps exhibiting a symmetric or anticipatory profile (LL region of Fig. \ref{fig:meanVSmin}). Those co-jumps are usually $S=2$ stocks co-jumps (76\%), but their size can go up to $S=87$ stocks. 

Hence, the most striking conclusion of this section is that many large co-jumps are in fact explained by endogenous dynamics and propagate across stocks, rather than being due to impactful external news. A (in)famous example of such propagation is the flash crash of May 6th 2010, where the S\&Pmini crashed in less than 30min, due to a sell algorithm set with an excessively high execution rate. This crash triggered a price drop in other US stocks. Here, our results suggest that this synchronization phenomenon is not such a rare event and actually happens quite often \cite{gerig2012high, bormetti2015modelling}. 


This finding is further supported by examining the correlation of the individual jump time-series composing a co-jump. 
Naively, one would expect large co-jumps to be exogenous, i.e. induced by news. As a result, the stocks involved in the co-jump should all share the same profile around the jump, as in Fig.~\ref{fig:market-cojump} for example. 
In fact, Fig.~\ref{fig:cojump-correlation} shows that there remain many co-jumps whose constituting univariate jump profiles are weakly correlated (see Appendix \ref{app:cojump-correlation} for more details). 
We also refer the reader to additional statistics on co-jumps in Appendix~\ref{app:complementary_result_cojumps}. 
For example, Fig. \ref{fig:sector_jump_proj} shows that the sector jumps are not all exogenous, as discussed in Section \ref{subsec:reflexivity}.










\section{Conclusion}

Thanks to an unsupervised approach based on wavelet scattering coefficients, we have identified three main directions along which price jumps can be classified. The first, well-known direction relates to the time-asymmetry of the volatility of the price around the jump and results in three classes of jumps, endogenous, exogenous and anticipatory. 

We also evidenced that mean-reversion and trend are important features for classification. This allowed us to identify three additional classes of jumps, ``mean-reverting'', ``trend-aligned'' and ``trend-anti-aligned'' which concerns a significant portion of the dataset. Thanks to this classification we have shown that a large portion of the jumps are endogenous or anticipatory jumps, confirming -- but also making much more precise -- the main conclusions of \cite{vol_news_jp, marcaccioli2022exogenous}. 

Extending our analysis to co-jumps, we have gathered several pieces of evidence that a large proportion of these co-jumps should also, quite surprisingly, be classified as endogenous in the sense that they seem to originate from the contagion of one single endogenous jump triggering the jump of possibly many others. One striking signature of such a scenario is the power-law distribution of co-jump sizes, which is indeed close to that predicted by a critical branching (contagion) process. Such a broad, power-law distribution of co-jump sizes was noted previously for different datasets in \cite{vol_news_jp, calcagnile2018collective, aubrun2023multivariate}. Further work should focus on higher frequency data that would allow one to dissect more precisely the contagion mechanism and ascertain that many large co-jumps are indeed {\it not} triggered by exogenous news, but related to the close-knit nature of financial markets that brings them close to critical fragility, as argued many times in the past, see e.g. \cite{bak1995complexity, bouchaud2011endogenous, fosset2020endogenous, moran2023temporal} and refs. therein. 

Unlike parametric fit of the time-series, the wavelet scattering embedding is defined and can be computed for any time-series. As such, our study could be transposed to other fields as well. 



\section*{Acknowledgements} \label{sec:acknowledgements}
We would like to thank Riccardo Marcaccioli for sharing his work on the detection of jumps in financial data and ideas on how to classify events based on their asymmetry. We are also very grateful to Guillaume Maitrier for helping us visualize and understand what happened in the Limit Order Book. We also thank Marcello Rambaldi and Iacopo Mastromatteo for their help with news data. Finally we would like to thank Maria Flora, Jérôme Garnier-Brun and Samy Lakhal for insightful discussions. 
 
This research was conducted within the
Econophysics \& Complex Systems Research Chair, under the aegis of the Fondation du Risque, the Fondation de l’Ecole polytechnique, the Ecole polytechnique and Capital Fund Management.



\bibliographystyle{vancouver}
\bibliography{biblio}

\clearpage

\newpage
\appendix
\onecolumngrid

\section{Benchmark: Validation through synthetic}\label{app:benchmark}

\begin{figure}[h]
\centering
\includegraphics[width=\textwidth]{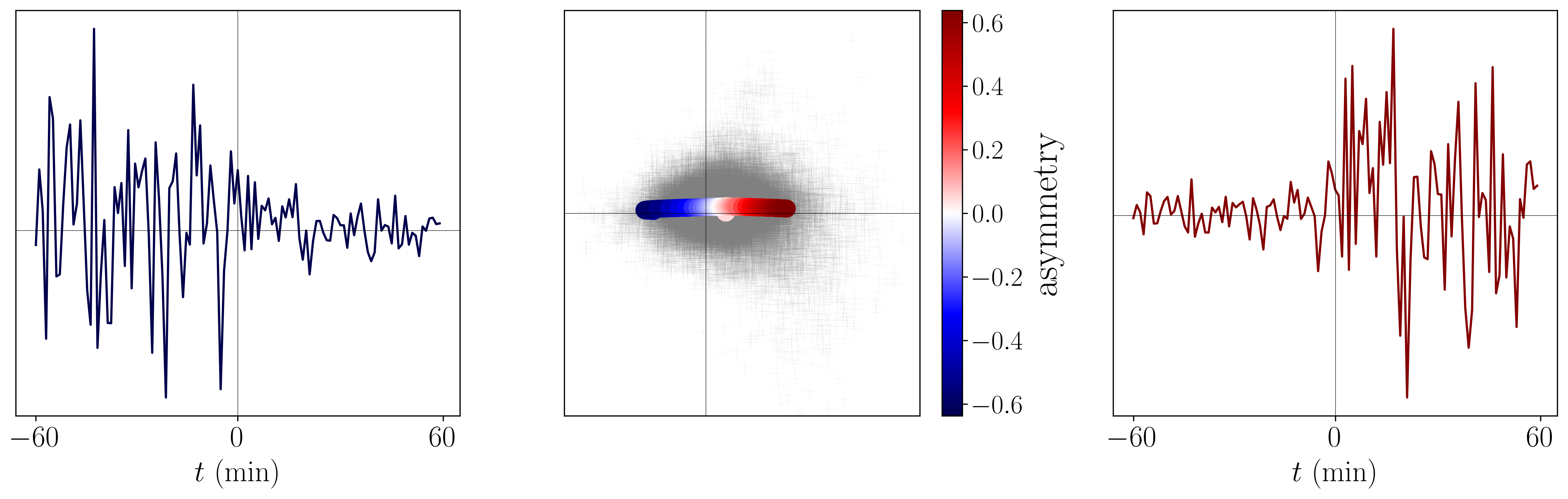}
\caption{
Middle: Projection of our dataset on the reflexive direction $D_1$ (horizontal axis) and mean-reverting direction $\widetilde{D}_2$ (vertical axis).
Benchmark jumps projection moves from left (anticipatory) to right (exogeneous). Left and right figures show two extreme benchmark time-series.
}
\label{fig:benchmark}
\end{figure}
In order to verify that the $D_1$ direction indeed measures reflexivity,
we create synthetic time-series with volatility profiles of varying time-asymmetry and apply our classification. 
Relying on \cite{marcaccioli2022exogenous}, we construct jump time-series using the power law representation of Equation~\eqref{eq:synthetic_jump_benchmarck}, with $t_c=-0.5$min, and $d=0.5$. We adjust the parameters $(N_r,N_l,p_r,p_l)$ to render the asymmetry of the signal. The time-series are then multiplied by a Gaussian noise (the same noise for all time-series). 
We then compute the features $\Phi(x)$ of each time-series $x$ and project it on the 2D space formed by our time-asymmetry and mean-reversion directions.
Fig.~\ref{fig:benchmark} shows these projections, the color code corresponds to the asymmetry parameter.
It clearly appears that the $D_1$ direction measures the time-asymmetry of the volatility. 

\clearpage
\section{Distribution of reflexivity/mean-reversion/trend scores}\label{app:dis_j1}

\begin{figure}[h!]
\centering
\begin{subfigure}[b]{0.3\textwidth}
    \centering
    \includegraphics[width=\linewidth]{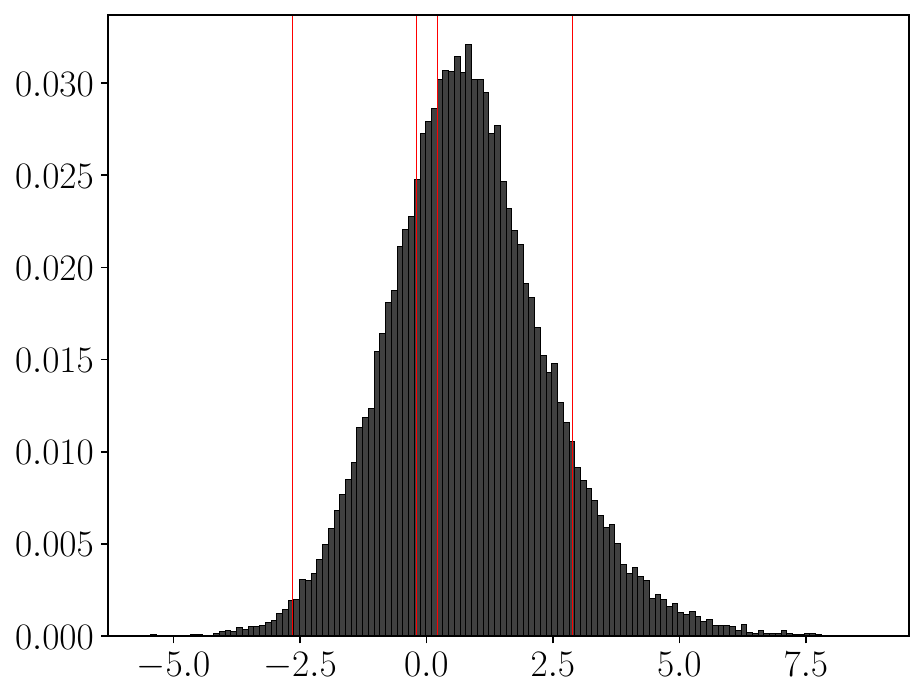}
    \caption[]{Reflexivity}    
\end{subfigure}
\begin{subfigure}[b]{0.3\textwidth}
    \centering
    \includegraphics[width=\textwidth]{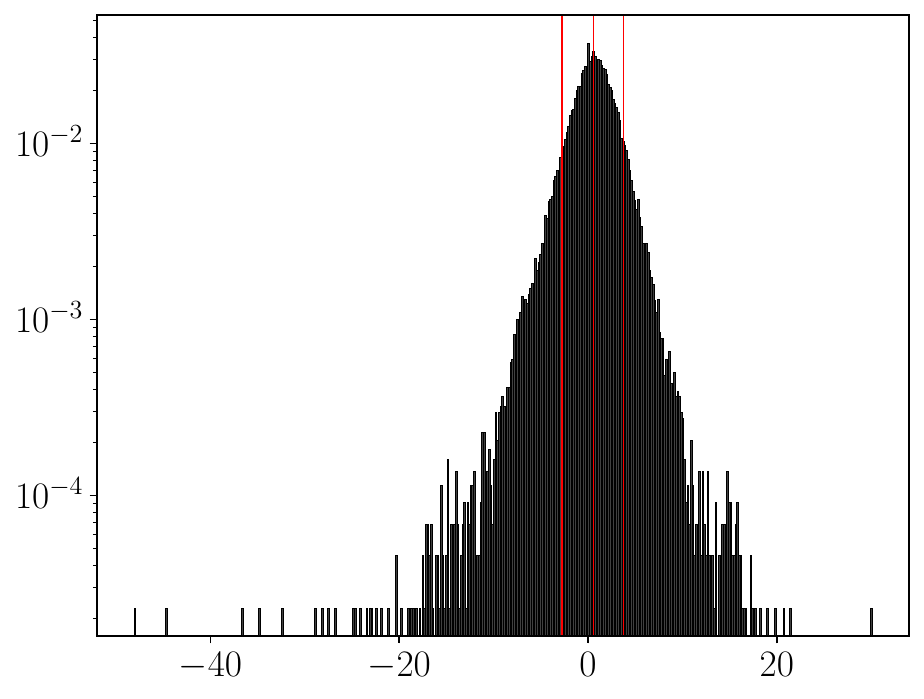}
    \caption[]{Mean-reversion}
\end{subfigure}
\begin{subfigure}[b]{0.3\textwidth}
    \centering
    \includegraphics[width=\textwidth]{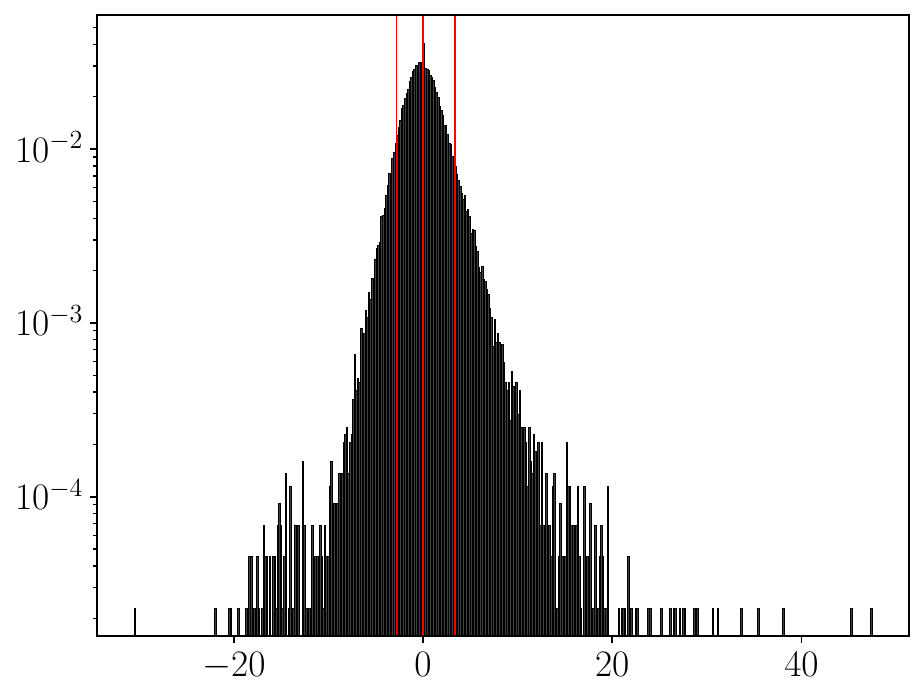}
    \caption[]{Trend}
\end{subfigure}
\caption{
Distribution of reflexivity $D_1(x)$, mean-reversion $\widetilde{D}_2$ and trend $\widetilde{D}_3$ scores used in this paper to identify classes of jumps.
The red vertical lines indicate the quantiles used to delimit the zones for the jumps taken into account when computing the average profiles in Figs.~\ref{fig:proj_on_x_profiles},\ref{fig:mr-projection},\ref{fig:td-projection}.
}
\label{fig:imagW_dis}
\end{figure}

\begin{figure}[h!]
\centering
\includegraphics[width=0.4\linewidth]{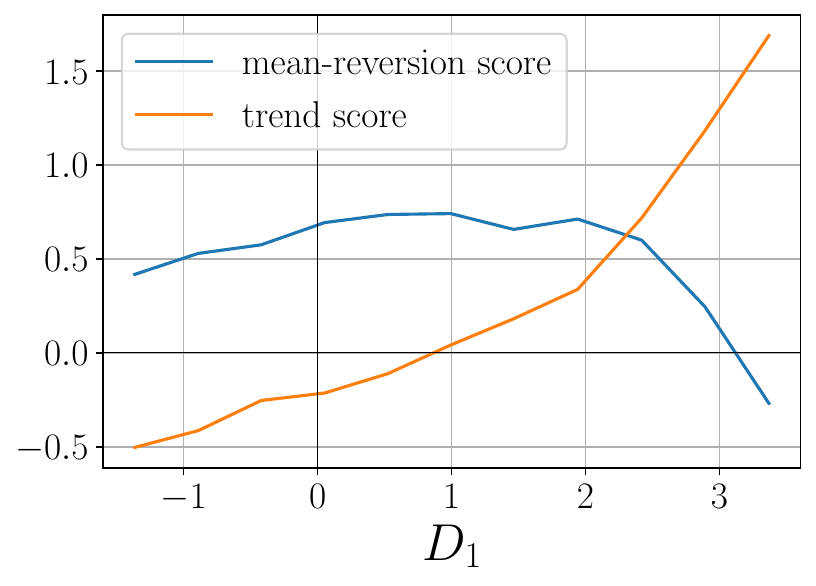}
\caption{
Mean-reversion and trend scores along reflexivity of a jump.
}
\label{fig:corr-refl-mr-td}
\end{figure}

\newpage
\section{Projection grid}

\subsection{Signed projections}\label{app:signed_grid}
\subsubsection{asymmetry VS mean-reversion}\label{app:signed_grid_mr}

\begin{figure}[h!]
    \centering
    \includegraphics[width = \linewidth]{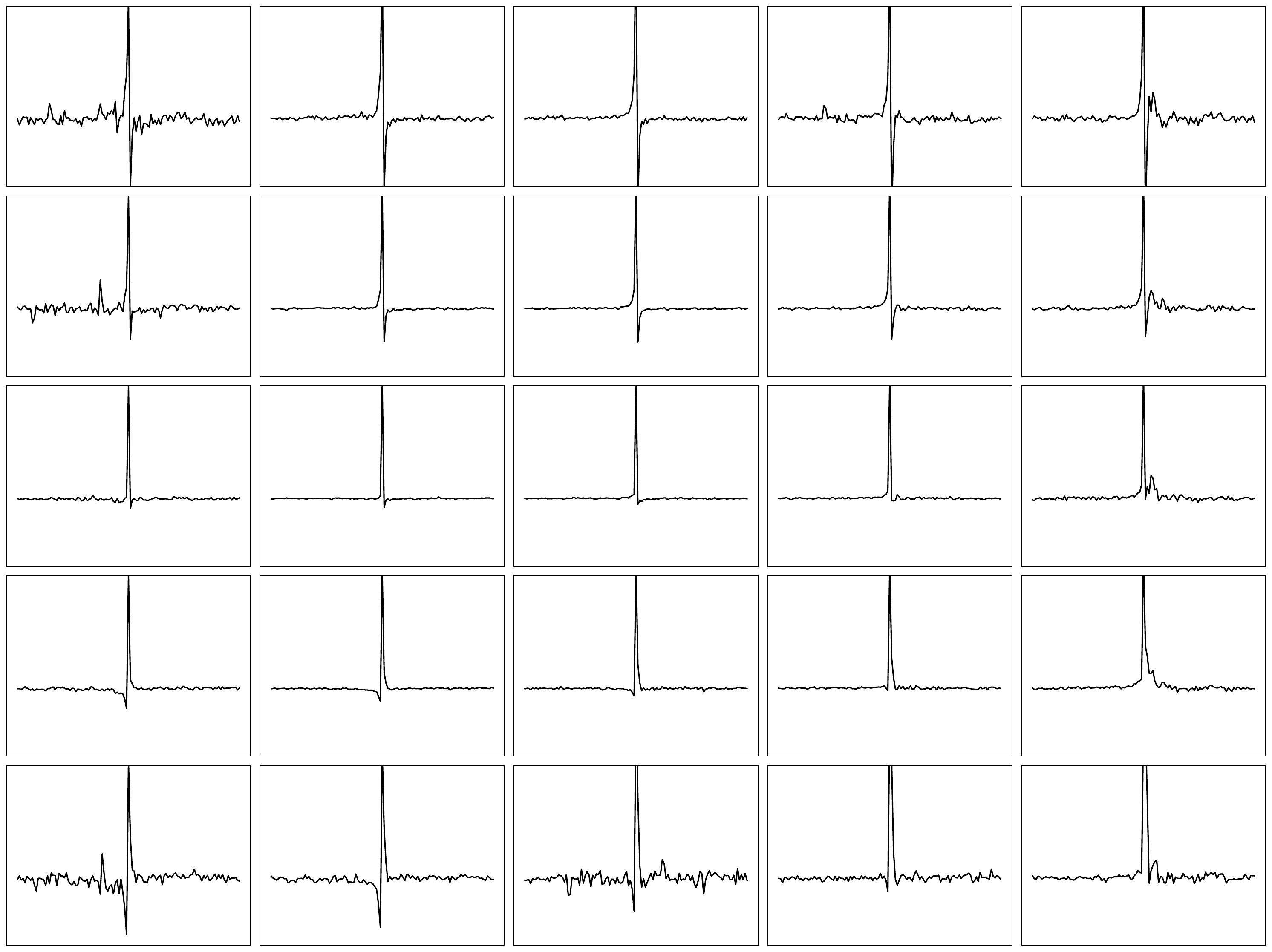}
    \caption{
    Average jump-aligned return profiles $\nosx(t)=\text{sign}(x(0))\,x(t)$. 
    Each plot represents the average over the jumps whose 2D projection falls in the respective box in the upper figure in Fig.~\ref{fig:proj_qu}.
    }
    \label{fig:signed_grid}
\end{figure}
\clearpage

\subsubsection{asymmetry VS trend}\label{app:signed_grid_tr}
\begin{figure}[h!]
    \centering
    \includegraphics[width = \linewidth]{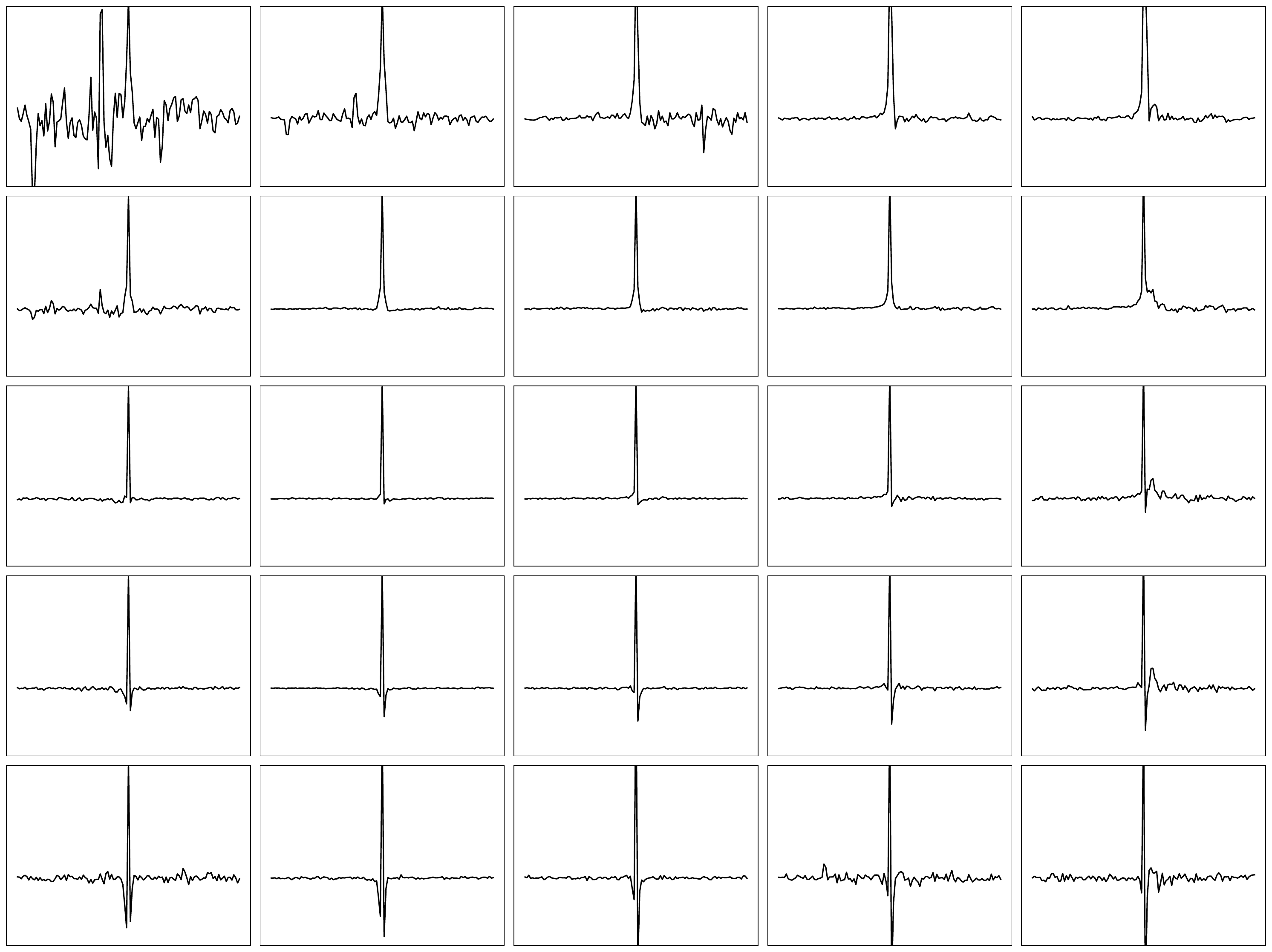}
    \caption{Average jump-aligned return profiles $\nosx(t)=\text{sign}(x(0))\,x(t)$. 
    Each plot represents the average over the jumps whose 2D projection falls in the respective box in the lower figure of Fig.~\ref{fig:proj_qu}.}
    \label{fig:signed_grid}
\end{figure}
\clearpage

\subsection{absolute projection - Asymmetry VS mean-reversion}\label{app:abs_grid}

\begin{figure}[h!]
    \centering
    \includegraphics[width = \linewidth]{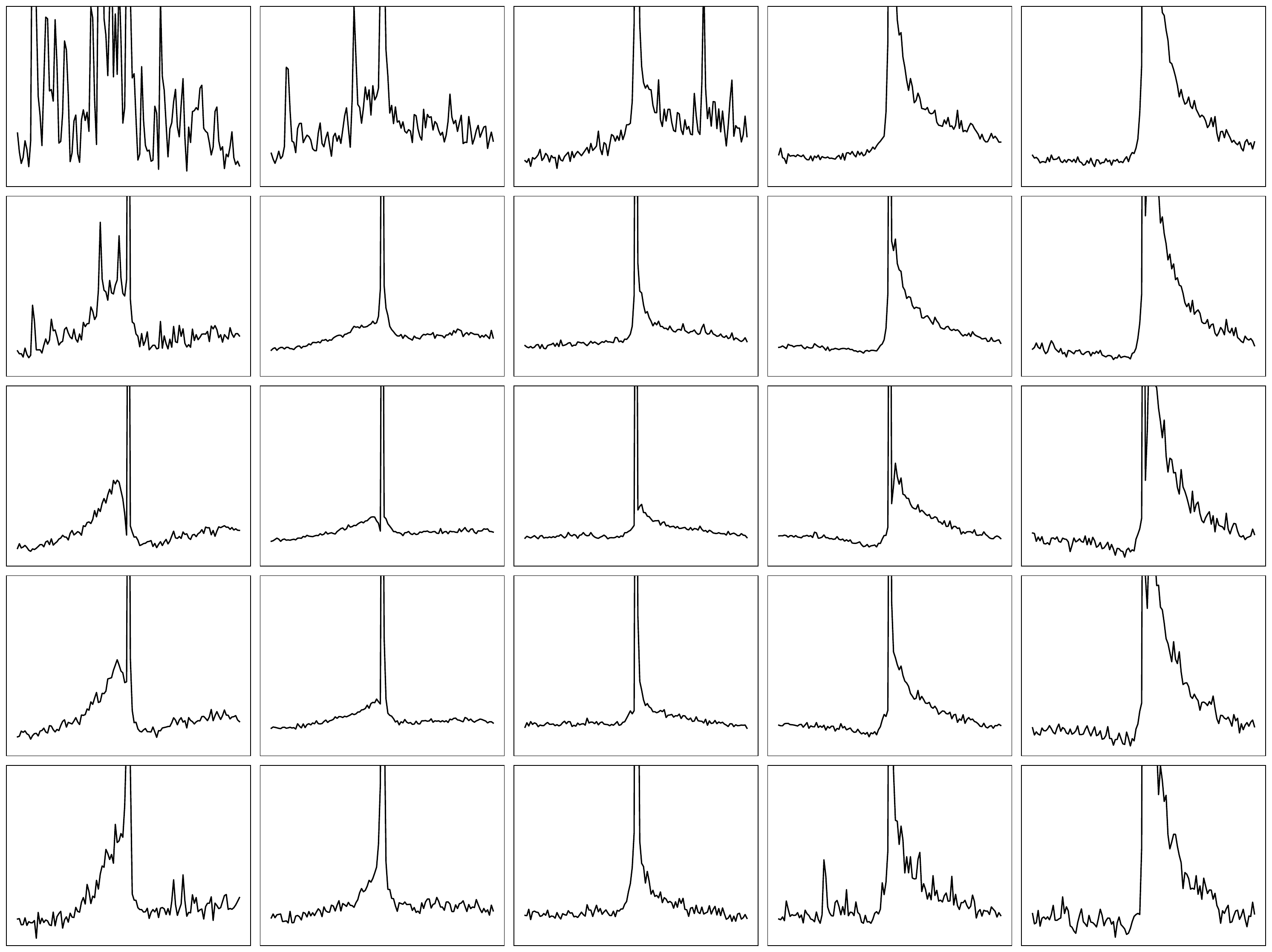}
    \caption{Average absolute profiles $|x(t)|$. Each plot represents the average over the jumps whose 2D projection falls in the respective box in the upper figure of Fig.~\ref{fig:proj_qu}.}
    \label{fig:abs_grid}
\end{figure}

\clearpage
\section{Example from the LOB}\label{app:LOBexample}

\begin{figure*}[h!]
\centering
\begin{subfigure}[b]{0.23\textwidth}
    \centering
    \includegraphics[width=\textwidth]{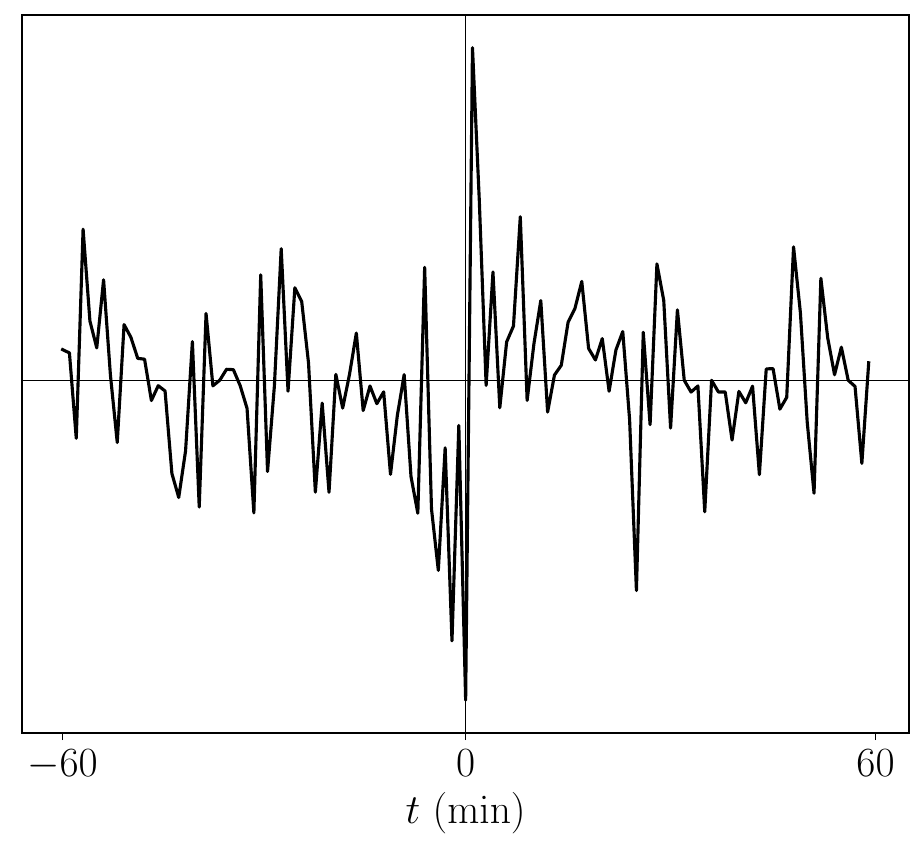}
    \caption[]%
    {{\small Jump-aligned return profile $\nosx(t)=\text{sign}(x(0))\,x(t)$ of the jump which occurred on stock HCA (Hospital Corporation of America), on the 2017-3-21 at 11h37. mean-reverting score: 2.61.}}    
    \label{fig:mean and std of net14}
\end{subfigure}
\hfill
\begin{subfigure}[b]{0.75\textwidth}  
    \centering 
    \includegraphics[width=\textwidth]{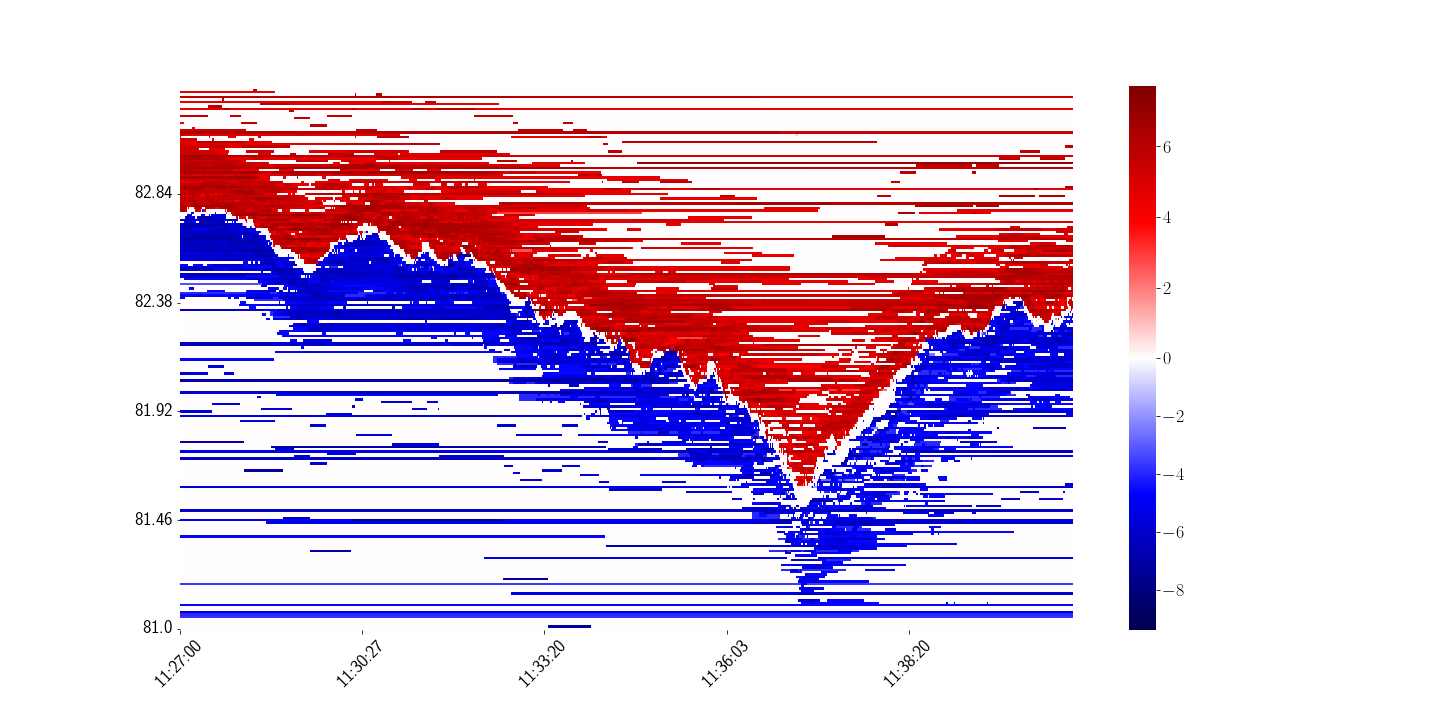}
    \caption[]%
    {{\small Respective LOB illustration. Each colored square represents an order in the LOB whose price is referred on the $y$-axis. The $x$-axis describes the time. Red is for the ask side, blue for the bid side. The color bar depicts the size of the order and is in log scale.}}    
    \label{fig:mean and std of net24}
\end{subfigure}
\vskip\baselineskip
\begin{subfigure}[b]{0.23\textwidth}  
    \centering 
    \includegraphics[width=\textwidth]{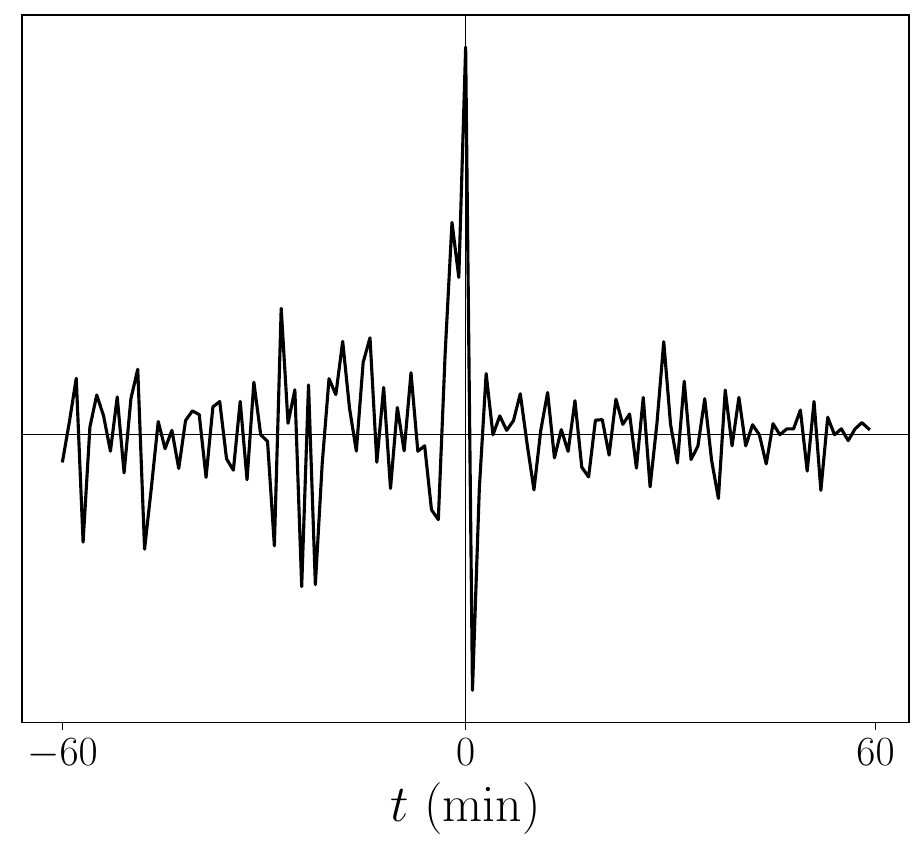}
    \caption[]%
    {{\small Jump-aligned return profile $\nosx(t)=\text{sign}(x(0))\,x(t)$ of the jump which occurred on stock INCY (Incyte), on the 2021-10-14 at 11h00. Mean-reverting score: 3.25.}}    
    \label{fig:mean and std of net34}
\end{subfigure}
\hfill
\begin{subfigure}[b]{0.75\textwidth}
    \centering 
    \includegraphics[width=\textwidth]{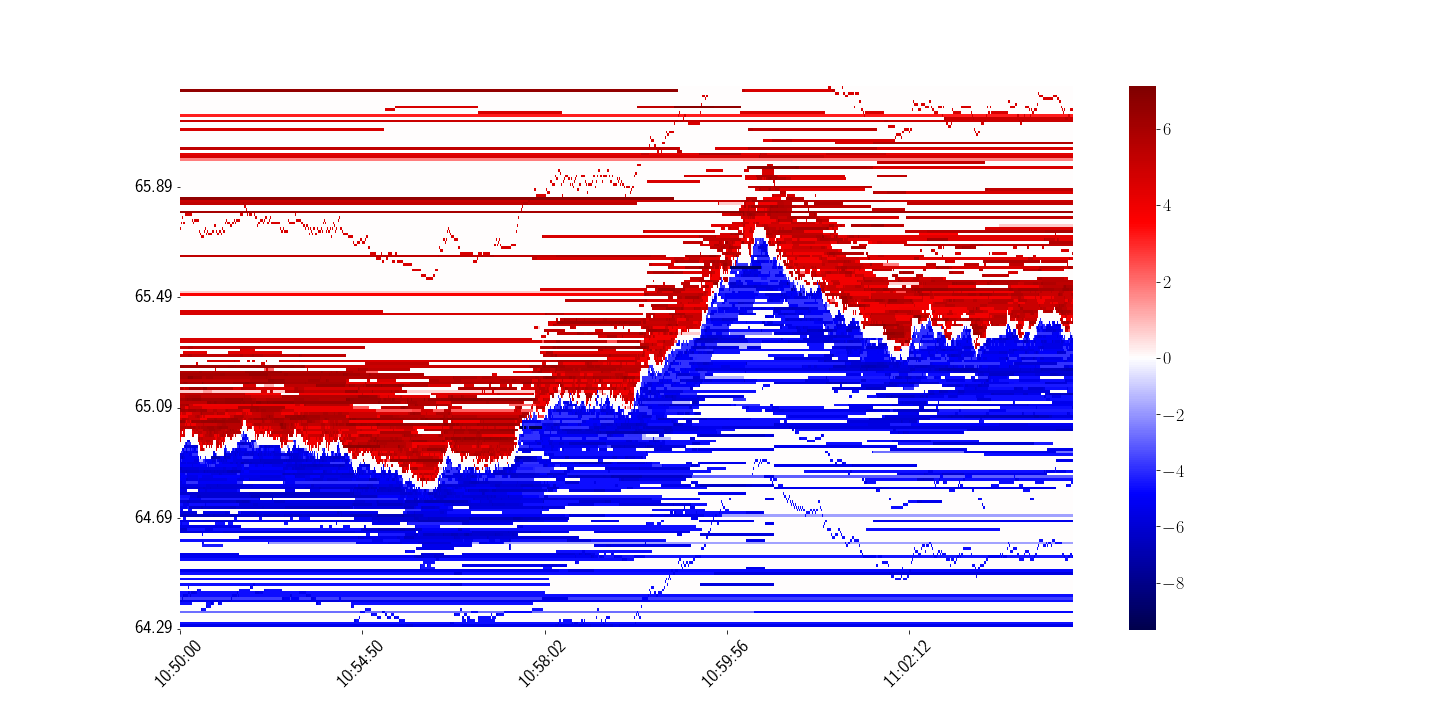}
    \caption[]%
    {{\small Respective LOB illustration. Each colored square represents an order in the LOB whose price is referred on the $y$-axis. The $x$-axis describes the time. Red is for the ask side, blue for the bid side. The color bar depicts the size of the order and is in log scale.}}    
    \label{fig:mean and std of net44}
\end{subfigure}
\caption[ The average and standard deviation of critical parameters ]
{\small Limit Order Book illustration of two strongly mean reverting jumps.} 
\label{fig:LOBillustration}
\end{figure*}

\clearpage
\section{Heterogeneous Near-Critical Branching Processes}\label{app:branching}

Consider a simple branching process where a single event can trigger on average $\varphi$ new events. It is well known that when $\varphi\to 1$, very large ``avalanches'' of events can occur. In fact, the probability that an avalanche of total size $S$ is triggered by a single event takes, in the limit $\varepsilon = 1 - \varphi \to 0$, the following scaling form:
\begin{equation} 
    P(S |\varepsilon) \propto S^{-3/2} e^{-\varepsilon^2 S}.
\end{equation}
Now suppose that the proximity to the critical point $\varphi = 1$ is itself random, reflecting the time varying fragility of the market and/or the intrinsic propensity of a shock to propagate across stocks. We will assume for simplicity that $\varepsilon$ has uniform distribution between $\varepsilon_{\min}$ and $1$. The observed distribution of avalanche sizes (in our case co-jump sizes) is then given by the following mixture:
\begin{equation} 
    P(S) = \frac{1}{1 -\varepsilon_{\min} } \int_{\varepsilon_{\min}}^1 {\rm d}\varepsilon \, P(S |\varepsilon) \propto S^{-3/2} \int_{\varepsilon_{\min}}^1 {\rm d}\varepsilon \,  e^{-\varepsilon^2 S}.
\end{equation}
After a change of variable, 
the integral over $\varepsilon$ can be rewritten as
\[ 
\int_{\varepsilon_{\min}}^1 {\rm d}\varepsilon \,  e^{-\varepsilon^2 S} = \frac{1}{\sqrt{S}} \int_{\varepsilon_{\min} \sqrt{S}}^{\sqrt{S}} {\rm d}u \,  e^{-u^2}
\]
Now, in a intermediate regime where $S \gg 1$ but 
$\varepsilon_{\min} \sqrt{S} \lesssim 1$, the  integral is close to $\sqrt{\pi}/2$, and one finally finds 
\begin{equation} 
    P(S) \propto S^{-1 - \tau} F(\varepsilon_{\min} \sqrt{S}), \qquad \tau=1,
\end{equation}
with $F(x)$ decreasing fast as $x \uparrow$. Hence this simple model predicts $\tau=1$ (i.e. a Zipf law) for co-jump sizes, truncated beyond $S \sim \varepsilon_{\min}^{-2}$. From the data shown in Fig \ref{fig:co-jumps-stats_0}, we estimate $\varepsilon_{\min} \sim 0.1$. In other words, the market does not have to be poised extremely close to criticality to explain a broad power-tail for the co-jump size distribution. 

Note finally that we could relax the hypothesis that the distribution of $\varepsilon$ is strictly uniform. In the scaling regime, one only needs this distribution to be constant in the vicinity of $\varepsilon_{\min}$. The calculation above can be extended to cases where 
the distribution of $\varepsilon$ is of a power law type close to zero, i.e. behaves as $\varepsilon^\gamma$ for $\varepsilon \to 0$. In this case, one finds $P(S) \propto S^{-1 - \tau}$ with $\tau = 1 + \gamma/2$.

\newpage

\section{Statistics on co-jumps}\label{app:complementary_result_cojumps}

\begin{figure}[h!]
\centering
\begin{subfigure}[t]{0.45\textwidth}
    \centering
    \includegraphics[width=\textwidth]{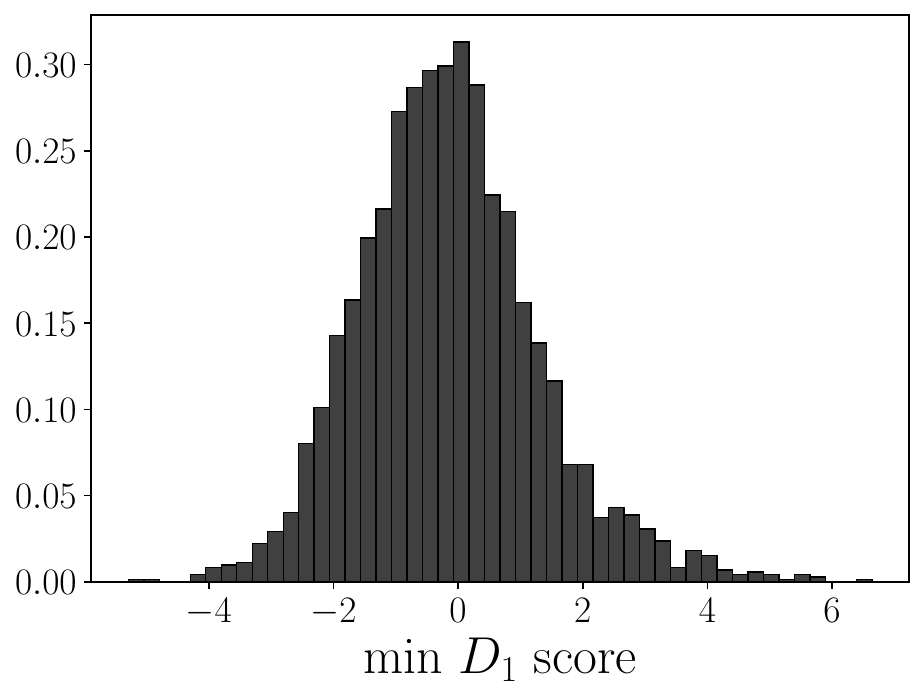}
    \caption{
    Distribution of the $D_1$ score of the most endogenous jumps of each co-jump (the leftmost jump in our 2D projections (see Fig. \ref{fig:proj_qu}) of all jumps belonging to a same co-jumps).
    }
    \label{fig:distribution_min_x}
\end{subfigure}
\hfill
\begin{subfigure}[t]{0.45\textwidth}
    \centering
    \includegraphics[width=\textwidth]{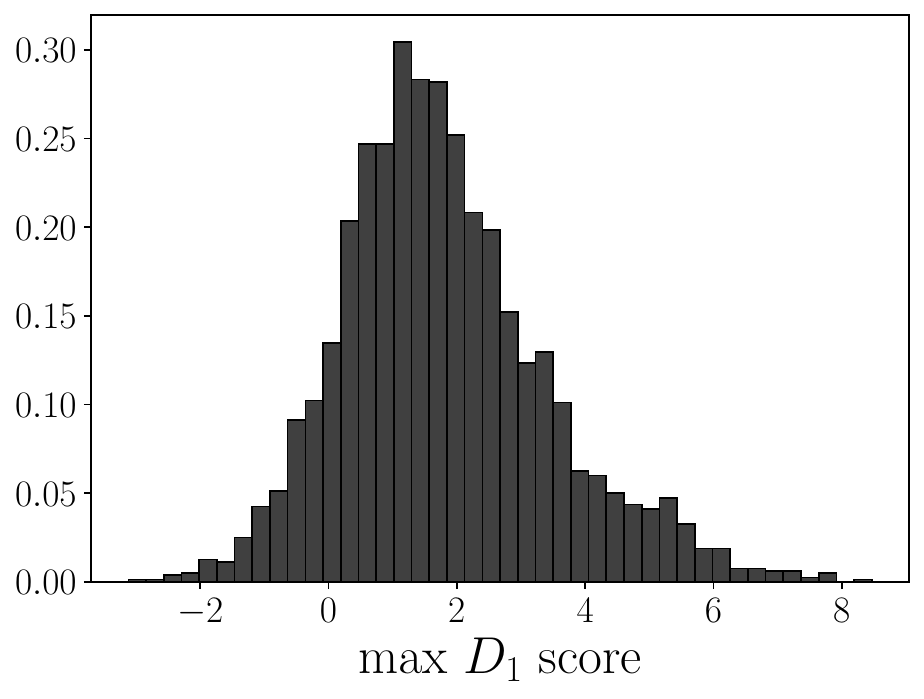}
    \caption{
    Distribution of the $D_1$ score  of the most exogenous jumps of each co-jump (the rightmost jump in our 2D projections (see Fig. \ref{fig:proj_qu}) of all jumps belonging to a same co-jumps).
    }
    \label{fig:distribution_max_x}
\end{subfigure}
\caption{
Statistics on co-jumps from the reflexive direction on co-jumps.
}
\label{fig:co-jump-stat}
\end{figure}


\begin{figure}
\centering
\begin{subfigure}[t]{0.45\textwidth}
    \centering
    \includegraphics[width=\textwidth]{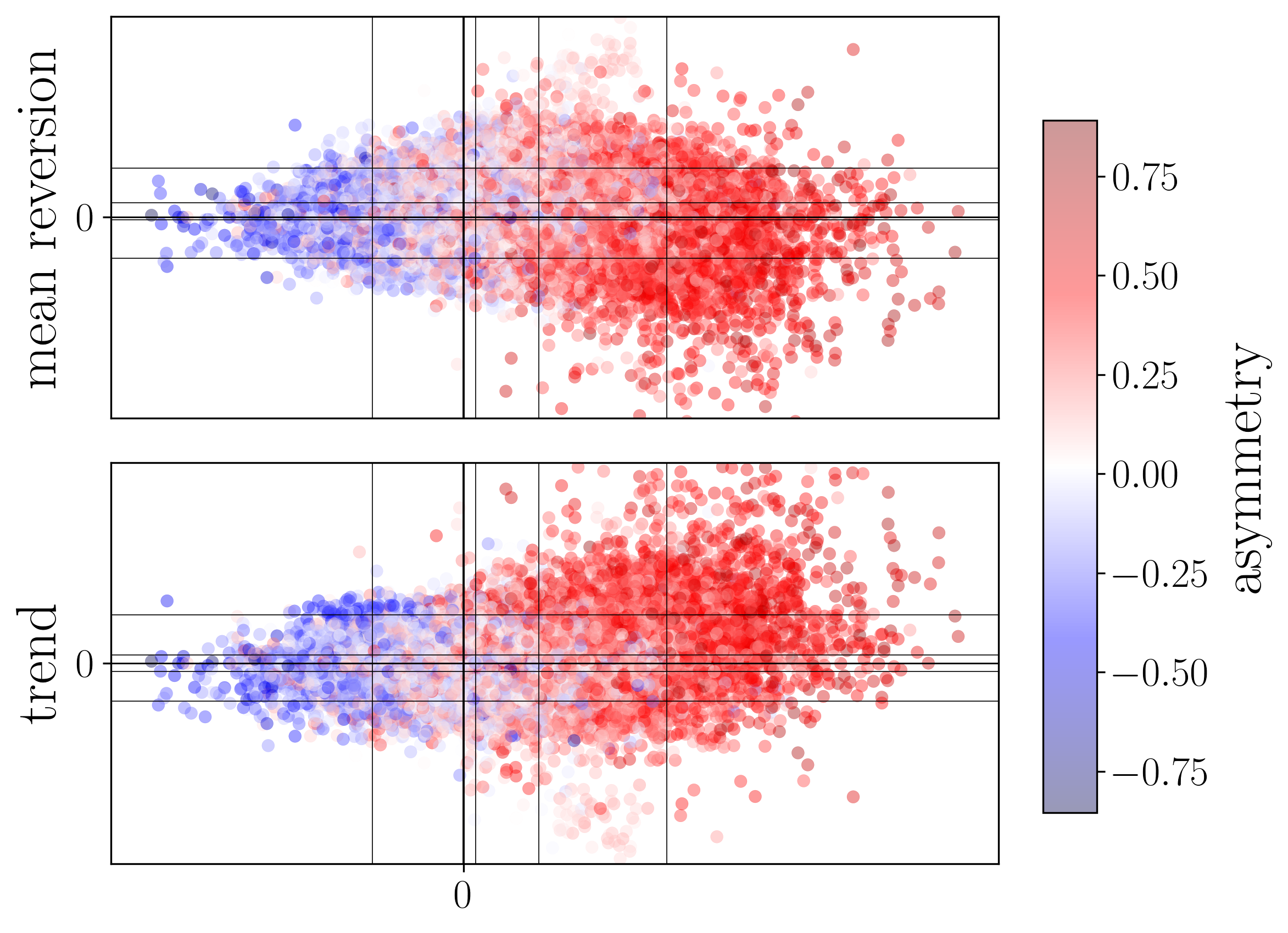}
    \caption{The 2D projections where the color represents the asymmetry of the jump computed with Eq.~\eqref{eq:asymmetry_def}.}
    \label{fig:projection_asym}
\end{subfigure}
\begin{subfigure}[t]{0.45\textwidth}
\centering
    \includegraphics[width=0.79\textwidth]{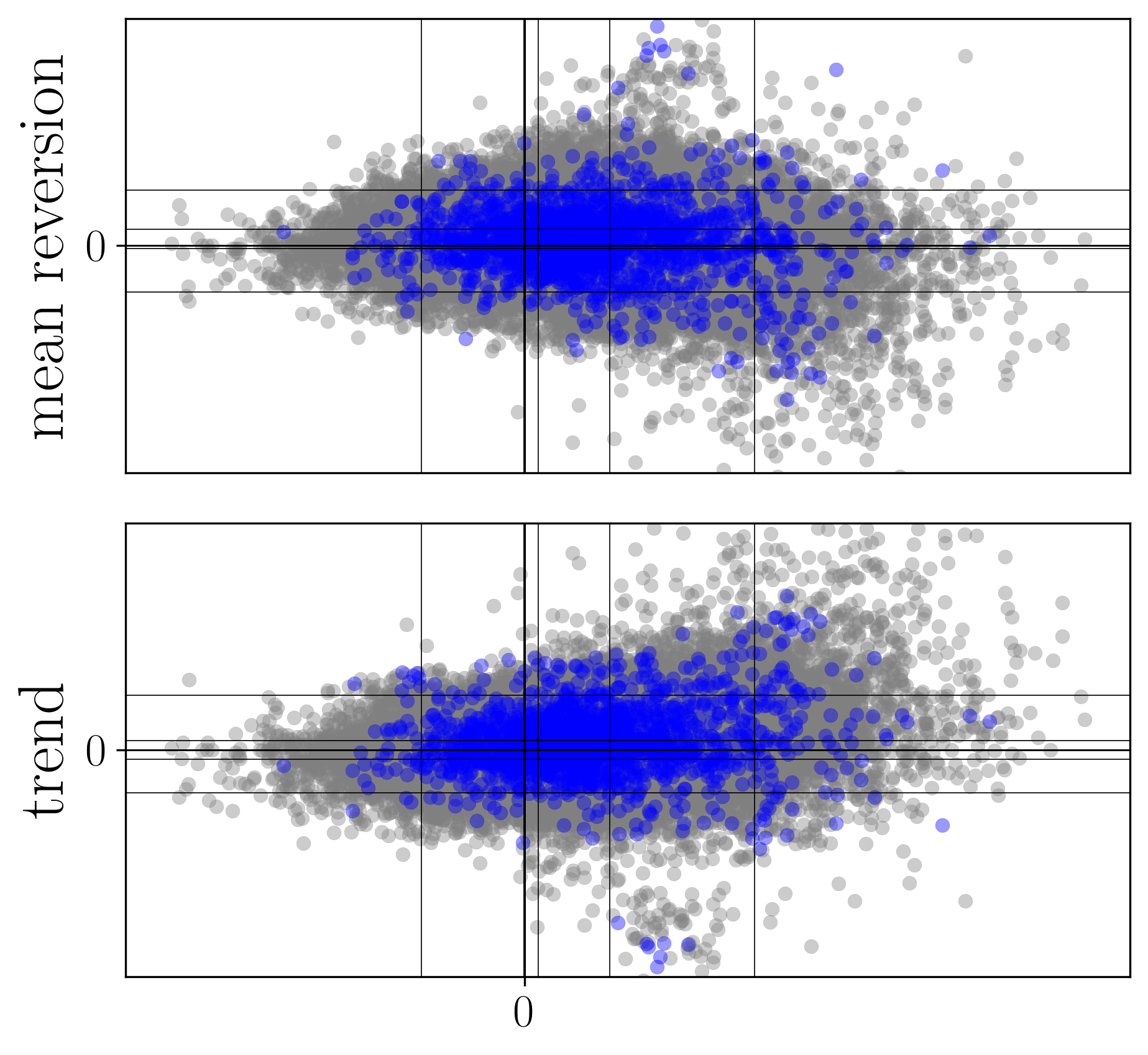}
    \caption{
    Projection of sector jumps in our 2D projections. The gray points represent the 2D projections of individual jumps of individual stocks. 
    The ``sector jumps'' are obtained by averaging time-series across socks of a same sector. The blue points represent the projection of those sector jumps time-series.
    }
    \label{fig:sector_jump_proj}
\end{subfigure}
\caption{Statistics on co-jumps from the reflexive direction}
\label{fig:co-jumps-stats}
\end{figure}


\begin{figure}
\centering
\includegraphics[width=0.5\linewidth]{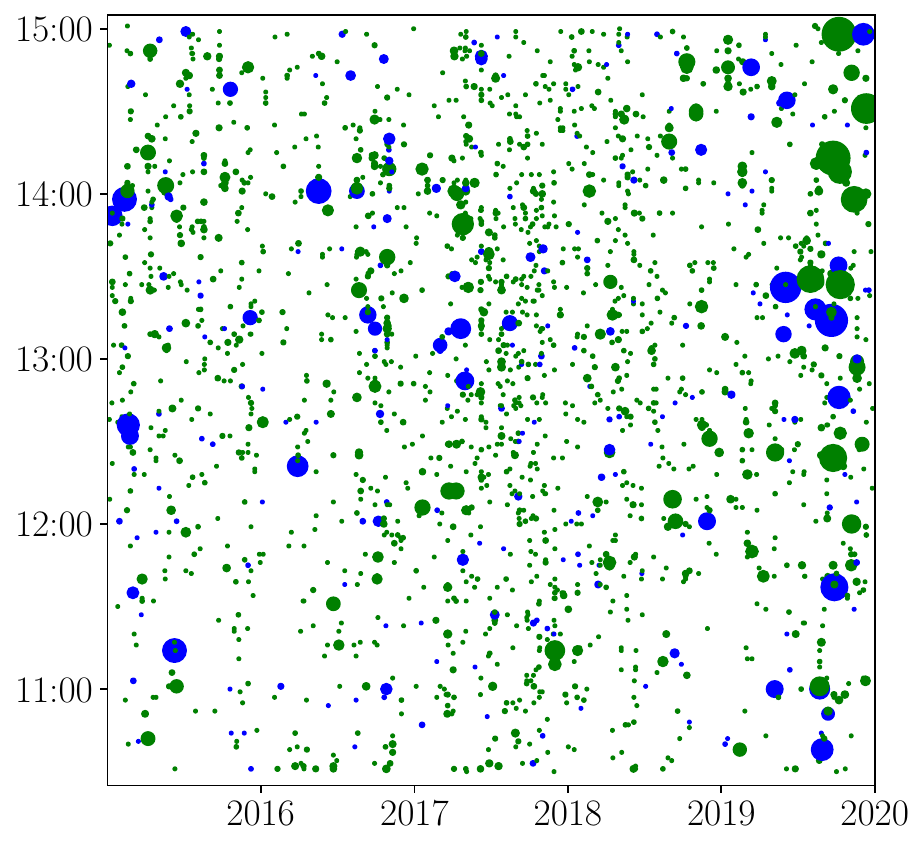}
\caption{
Visualization of our co-jump dataset (295 US stocks over 8 years). The
horizontal axis corresponds to the day in the sample and the vertical axis gives the time of day. The size of the circle
encodes the number of stocks simultaneously jumping in a given minute (see color bar). 
The circle's color signifies whether a co-jump coincided with a news related to one of the involved stocks within a 3-minute window (blue), or in the absence of reported news (green).
}
\label{fig:cojump_labels}
\end{figure}

\clearpage
\newpage

\section{Correlation of jump profiles in a co-jump}
\label{app:cojump-correlation}

In this appendix, we investigate to which extent the different price profiles in a co-jump are correlated to each other. 
To achieve this, we consider the average correlation of the trend score Eq.~\eqref{eq:trend} among the jumps in a given co-jump of size $S$, defined as 
\[
\rho = \frac{\sum_{k \neq k'=1}^S \widetilde{D}_3(x_k) \widetilde{D}_3(x_{k'})}{(S-1) \sum_{k=1}^S [\widetilde{D}_3(x_k)]^2 }.
\]

%
\begin{wrapfigure}{r}{0.5\textwidth}
\centering\includegraphics[width=0.5\textwidth]{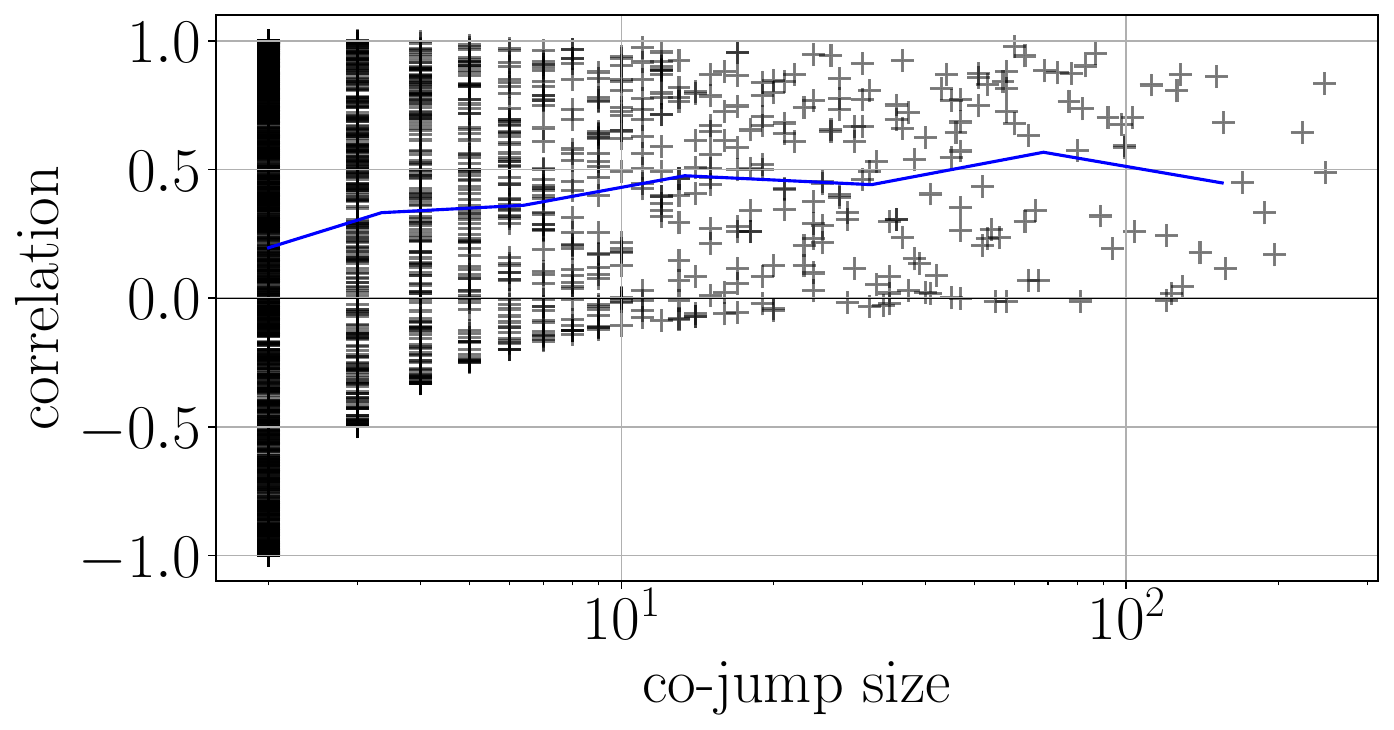}
\caption{
Correlation measure $\rho$ of a co-jump. 
Blue line represents average over bins of co-jumps of roughly the same size. 
As expected, the larger the co-jump the more correlated. Surprisingly, there remain weakly correlated ($\rho\approx0$) large co-jumps. 
}
\label{fig:cojump-correlation}
\end{wrapfigure}
\begin{figure}[b]
\centering
\begin{subfigure}{0.4\linewidth}
    \centering
    \includegraphics[width=0.7\linewidth]{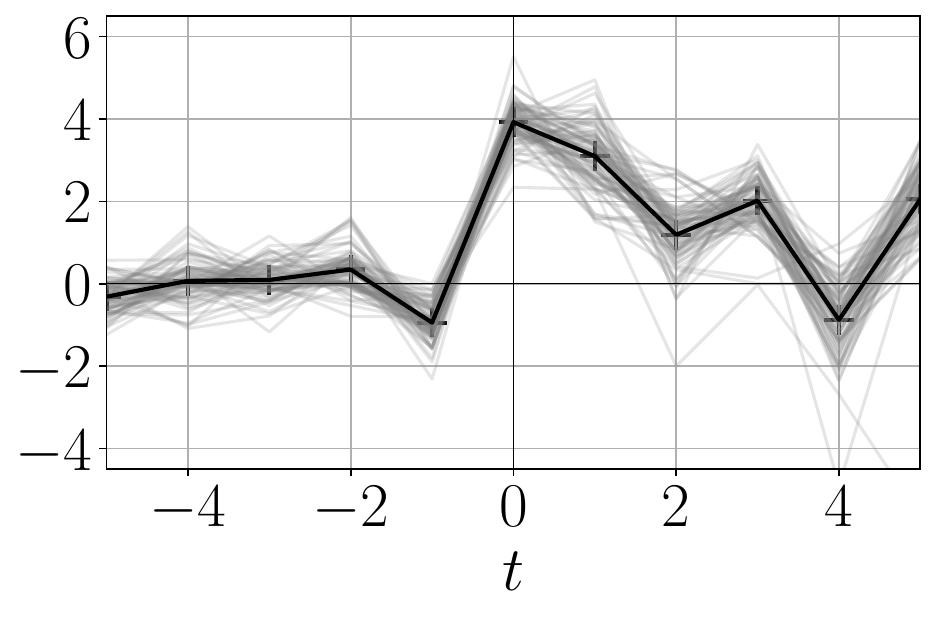}
    \caption{Strongly correlated}
    \label{fig:market-cojump}
\end{subfigure}%
\begin{subfigure}{0.4\linewidth}
    \centering
    \includegraphics[width=0.7\linewidth]{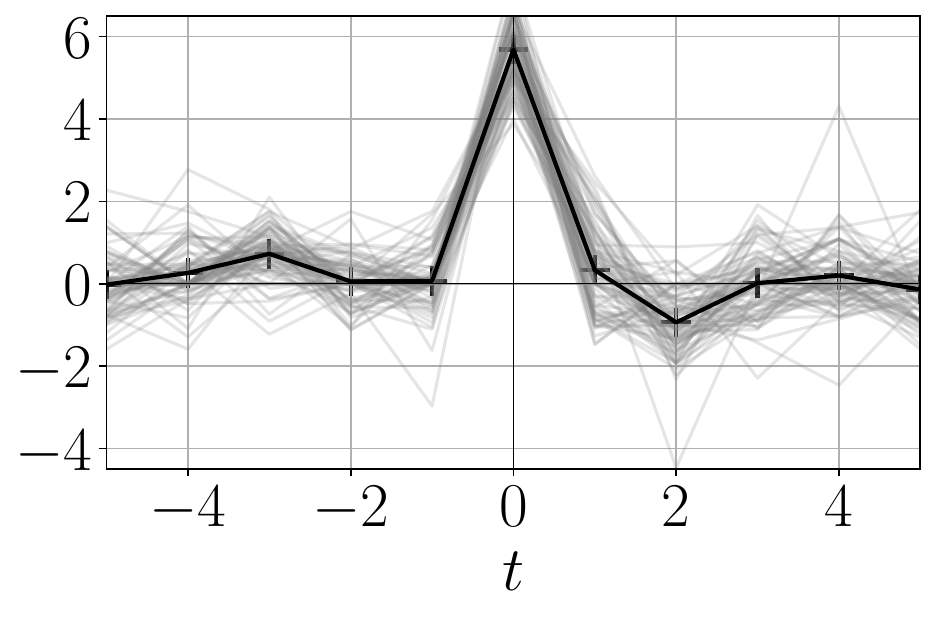}
    \caption{Weakly correlated}
    \label{fig:uncorr-cojump}
\end{subfigure}%
\caption{
Average profile of two co-jumps. 
The average (black curve) is taken over the profiles of the jumps involved in each co-jump (gray curves).
Left: a strongly correlated co-jump that exhibits a non-zero average profile. Each jump time-series in the co-jump is a variation around this average profile.
Right: a weakly correlated co-jump which has no meaningful average. 
The co-jumps are of size $83$ and $81$ respectively.
}
\end{figure}

%
Fig.~\ref{fig:cojump-correlation} shows that the larger the co-jump, the more correlated are its  constituents, although the effect is weak.
Jumps affecting the market in its entirety are more likely to have a common external reason (exogenous) and lead to the same profile. 
Strongly correlated co-jumps come down to a single jump time-series which can be accessed through the average of normalized jumps:
\[
\Av{}{\overline{x}}(t) = \Av{k}{\sigma_{k}^{-1}\overline{x}_k(t)}
\]
where $\sigma_{k} = \Av{t}{x^2_k(t)}^{\frac12}$.
Fig.~\ref{fig:market-cojump} shows an example of co-jump, of size $83$ with correlation $\rho=0.96$. 
We see that the average time-series $\Av{}{\overline{x}}$ is non-zero for $t\ne0$.

In line with the discussion of section \ref{sec:cojumps} about contagion-driven co-jumps, Fig.~\ref{fig:cojump-correlation} shows that there persist large co-jumps whose average correlation is close to zero, i.e. 
%
co-jumps composed of return time-series that are weakly correlated and that have no a priori reason to jump together, except through contagion. 
A typical example is shown on Fig.~\ref{fig:uncorr-cojump}, it is of size $81$ and correlation $\rho=0.06$. As we can see, averaging its different jumps makes small sense since the average $\Av{}{\overline{x}}$ is zero for $t\ne0$ (up to the variance). 
Far from being reduced to a single jump profile, such weakly large co-jumps could still be described by a small number of ``hidden'' profiles, depending on the ``dimensionality'' of the co-jump. Determining such dimensionality and hidden profiles would require applying a decomposition per co-jump, which would require more data on large co-jumps.


%
%


\newpage

\section{Additional figures}

\begin{figure*}[h]
\centering
\includegraphics[width=0.9\linewidth]{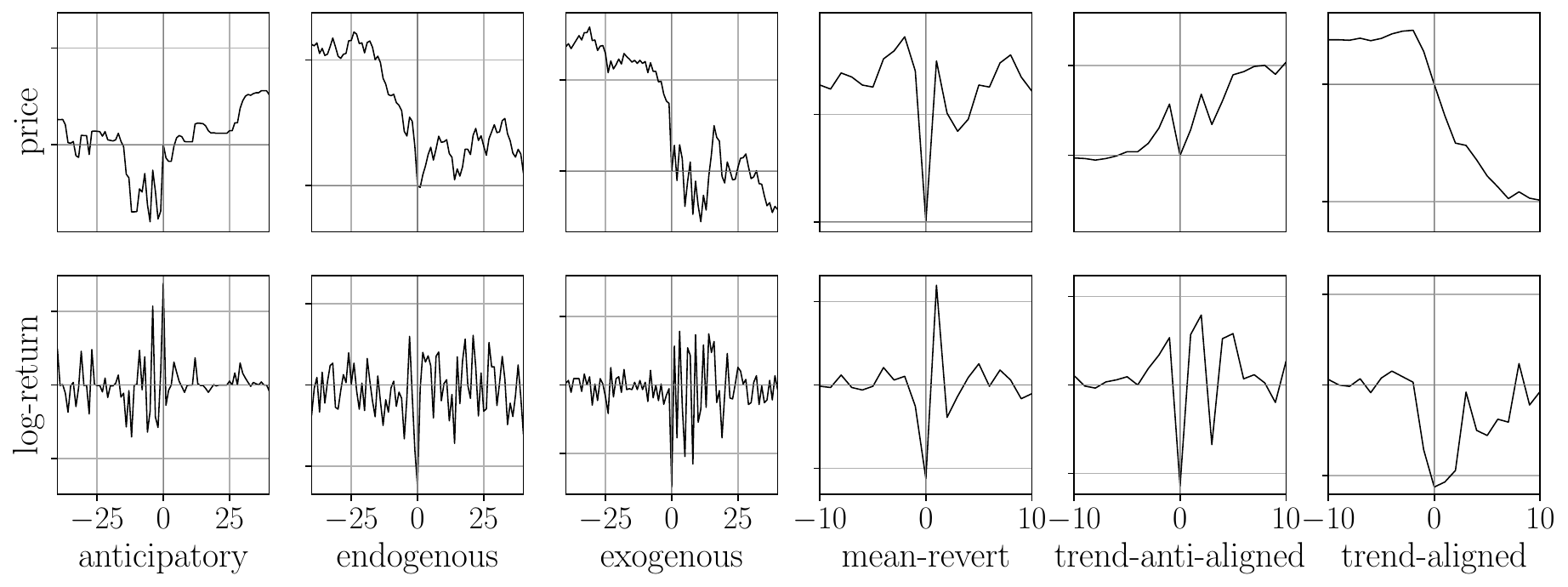}
\caption{
Classes of price jumps (observed examples). Each column shows an example of a class of jumps (price and log-return time-series).
The three first classes (anticipatory, endogenous, exogenous) are separated by measuring volatility asymmetry. The three last classes (mean-reverting, trend-anti-aligned, trend-aligned) are identified by analyzing the signed returns around the jump.
}
\label{fig:examples-observed}
\end{figure*}

 \begin{figure}
    \centering
    \includegraphics[width = 0.5\linewidth]{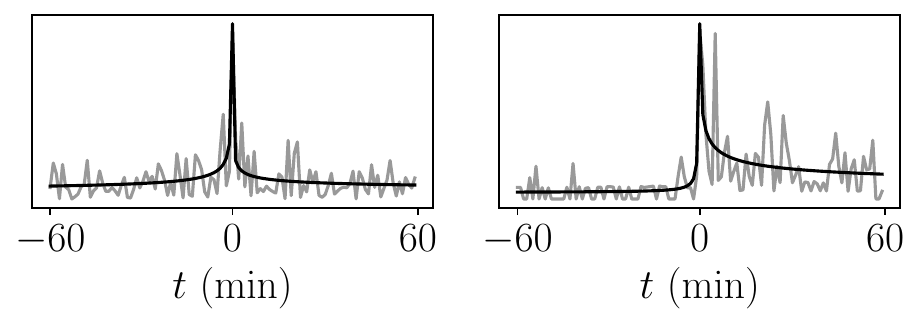}
    \caption{Examples of different asymmetry profiles in price jumps. Plain black lines are power law fits from \cite{marcaccioli2022exogenous} described in Eq.~\eqref{eq:synthetic_jump_benchmarck}. }
    \label{fig:sym_asym_examples}
\end{figure}

\end{document}